\documentclass[a4paper,11pt]{article}
\usepackage{jcappub} 
\usepackage{orcidlink}

\title{\boldmath Void probability function in the Quijote simulations}

\author{Snehasish~Bhattacharjee\,\orcidlink{0000-0002-XXXX-XXXX}}
\affiliation{Graduate Institute of Astronomy, National Central University,
300 Zhongda Road, 32001 Zhongli, Taiwan}

\emailAdd{snehasish@astro.ncu.edu.tw}

\abstract{The void probability function (VPF) is the likelihood of finding no tracers (galaxies or dark-matter haloes) within a given volume and is sensitive to the large-scale matter distribution. We investigate the VPF of dark matter haloes and galaxies using the \textsc{Quijote} N-body simulations across a range of cosmological models, including $\Lambda$CDM and several extensions. We examine its dependence on cosmology, redshift, halo mass, morphology, and sample dilution, and compare the results with theoretical predictions from the geometric hierarchical (GH) and negative binomial (NB) models. We find that the halo VPF is sensitive to cosmology and halo mass and follows the GH model at $z=0$, with $\Lambda$CDM showing the smallest deviation and the massive-neutrino cosmology the largest. At higher redshift and stronger sample dilution, the VPF departs from the GH model and approaches the NB model. The dependence on halo mass varies with cosmology, with low-mass haloes showing the largest deviations in $\Lambda$CDM, while intermediate- and high-mass haloes show stronger deviations in the modified-gravity scenario. In redshift space, the VPF shifts toward the NB model, with primordial non-Gaussianity producing the largest deviations. For galaxies, the VPF depends on morphology and cosmology, with ellipticals following the GH model and spirals following the NB model. We also find that cosmic variance introduces non-negligible scatter in the VPF measurements. Overall, these results suggest that the VPF could be useful as a computationally efficient probe to distinguish extensions of the standard $\Lambda$CDM model in upcoming large-volume galaxy surveys.
}

\begin{document}
\maketitle
\flushbottom

\section{Introduction}\label{sec:intro}

Cosmological observations suggest that we are living in a spatially flat, isotropic, and homogenous Universe undergoing a period of accelerated expansion \citep{reiss1998,perlmutter1999}. Over the years, a plethora of theoretical models have been proposed to explain this phenomenon, including modified gravity, quintessence, and dynamical dark energies \citep{Peebles2003}. Among these, the $\Lambda$CDM model emerged as the standard paradigm due to its simplicity and strong agreement with various observations \citep{frenk2012,bull2016}. However, the $\Lambda$CDM model postulates the Universe is composed of roughly $70\%$ dark energy and $25\%$ dark matter, neither of which has been directly observed. Hence, one of the most pressing questions in cosmology is to understand if dark energy and dark matter exist and to understand their nature.

The spatial distribution of galaxies on cosmological scales is sensitive to the properties of cosmological parameters, including dark energy, dark matter, neutrino masses, and primordial non-Gaussianity \citep{Deustua2002,Loverde2024,Ferraro2022}. Hence, a detailed analysis of the statistical properties of galaxies offers a robust framework for constraining these fundamental quantities and furthering our understanding of the underlying physical processes that drive the evolution and dynamics of the Universe.

Understanding the properties and distribution of voids in the galaxy distribution is important, since voids act as a key probe of nonlinear structure formation and capture statistical properties beyond the two-point correlation function \citep{cooray2002,sheth2004,padilla2005,fry2013,kurban2023}. A remarkable feature of voids is their hierarchical scaling behavior, which has been studied within the framework of the halo model \citep{fry2013} and rigorously tested in multiple redshift surveys \citep{Maurogordato1987,Vogeley1994,Fry1989,Maurogordato1992,Bouchet1993,croton2004,croton2007,conroy2005,tinker2008}. The Void Probability Function (VPF) is a statistical measurement of finding no galaxies (i.e., a void) within a given volume and is theoretically linked to all higher-order correlation functions of galaxy clustering \citep{White1979,Sharp1981,Fry1986,Balian1989,coles1991,Gaztanaga1993,Vogeley1994,Benson2003,baugh2004,croton2004b}. Thus, estimating the VPF offers a means of accessing information from all correlation orders while reducing computational expenses.

Various analytical models have been proposed to describe the VPF \citep{Fry1986}. The geometric hierarchical (GH) model, which assumes a hierarchical scaling of correlation functions, in which higher-order correlation functions are expressed in terms of the two-point correlation function with scale-dependent amplitudes, has been found to provide a good description of voids identified from dark matter halo distributions \citep{carruthers1983,Fry1986}. On the other hand, the negative binomial (NB) model, which is based on the negative-binomial distribution by generalizing the Poisson distribution to account for clustering-induced over-dispersion, results in a logarithmic form for the VPF and provides a good description of galaxy distributions \citep{hamilton1988,Elizalde1992,conroy2005,croton2004,fry2013,Hurtado-Gil2017,kurban2023}. However, several studies have observed deviations of the VPF from these models in certain scenarios. Refs \cite{croton2007,tinker2008} report that the redder galaxies in SDSS deviate from the NB model due to their stronger clustering compared to blue galaxies, while simulations by Refs \cite{Vogeley1994,lahav1993} indicate that the VPF of galaxies follows the NB model only in redshift space, and not in real space.

In contrast to previous studies  \cite{carruthers1983,Fry1986,hamilton1988,Elizalde1992,Vogeley1994,lahav1993,conroy2005,croton2007,tinker2008,kurban2023}, which have primarily focused on testing VPF within a single cosmological framework, this work presents a systematic investigation of the behaviour of the VPF across a range of extensions to the standard $\Lambda$CDM model. In particular, we consider variations in primordial non-Gaussianity, total neutrino mass, modified gravity, background density contrast, dark energy, and the standard $\Lambda$CDM parameters, using both dark-matter halo and mock galaxy catalogues obtained from the \textsc{Quijote} simulations. By varying one cosmological parameter at a time while keeping the others fixed allows us to isolate and compare the impact of individual parameters on the VPF in a robust way. The article is organized as follows: In Section~\ref{vsf}, we provide an overview of VPF. In Section~\ref{sim}, we describe the \textsc{Quijote} simulations and cosmological parameters used in this work. In Section~\ref{vpf:dm} and \ref{vpf:g}, we present our results for the dark matter haloes, and galaxies respectively, and in Section~\ref{con} we present our conclusions.

\section{Overview of Void Probability Function}\label{vsf}

The count probability distribution function \( P_N(R) \) is an important quantity that describes the probability of observing \( N \) galaxies within a volume of radius \( R \). VPF is a special case of this function, in which the likelihood of finding no galaxies \( P_0(R) \) has been reported to be connected with the hierarchy of all the higher-order correlation functions as follows \citep{White1979,Fry1986}:

\begin{equation}\label{eq1}
    P_{0}(R) = \exp \bigg\{\sum_{n=1}^{\infty}\frac{\big( - \overline{N} (R)  \big) ^{n}}{n !}\overline{\xi}_{n}(R)\bigg\},
\end{equation}
where $\overline{N} (R)$ represents the average number of galaxies within the volume of radius $R$ and $\overline{\xi}_{n}$ is the volume-averaged $n$-th order correlation function:
\begin{equation}
    \overline{\xi}_{n} = \frac{\int \xi_{n}dV}{\int dV}.
\end{equation}
Volume-averaged correlation functions follow a hierarchical pattern of the form 
\begin{equation}\label{eq2}
    \overline{\xi}_{n} = S_{n}\xi_{2}^{n-1},
\end{equation}
where the coefficients $S_{n}$ are scale independent and assume the form $S_{n} = n^{n-2}$ \citep{bernardeau1999}.  $ \xi_2(r)$ represents the two-point correlation function, defined as the excess probability ($dP$) of finding two galaxies separated by a distance $r$ within a volume $dV$ and can be written as \citep{peebles1980},
\begin{equation}
    dP = \bar{n} \,[1 + \xi_{2}(r)] \, dV,
\end{equation}
where $\bar{n}$ is the mean number density.
Using Eq. \ref{eq2} in  Eq. \ref{eq1} we obtain, 

\begin{equation}
      P_{0}(R) = \exp \bigg\{\sum_{n=1}^{\infty}\frac{\big( - \overline{N} (R)  \big) ^{n}}{n !}S_{n}\xi^{n-1}(R)\bigg\},
\end{equation}
where we write $\xi = \xi_{2}$ for simplicity. Taking the logarithm on both sides, we obtain the void statistic $\chi$ as follows:

\begin{equation}\label{eq3}
    \chi = \frac{-\log P_{0}}{\overline{N}} = \sum_{n=1}^{\infty}\frac{( -1 ) ^{n-1}}{n !}S_{n}(\overline{N}\xi)^{n-1} = \chi (\overline{N}\xi).
\end{equation}

This is the hierarchical scaling relation, which implies that the void statistic
$\chi = -\log P_{0}/\overline{N}$ depends solely on the scaling variable
$\overline{N}\xi$. The quantity $\overline{N}\xi$ combines information about both
the mean tracer density and the strength of clustering, and can be interpreted as
a measure of the degree of non-linearity within the sampled volume. $\overline{N}\xi \ll 1$ corresponds to large correlation length and/or low number densities, for which $P_{0} \simeq e^{-\overline{N}}$ and
$\chi \rightarrow 1$, recovering the Poisson limit. In contrast,
$\overline{N}\xi \gg 1$ arise on small scales and/or in high-density, strongly
clustered environments, where higher-order correlations become important. In this
regime, the behaviour of the VPF becomes a powerful tool to test various models of hierarchical clustering.

As mentioned earlier, the GH and the NB models provide a good fit for dark-matter haloes \cite{carruthers1983}, and for galaxies \citep{hamilton1988,Elizalde1992,conroy2005,croton2004,fry2013,Hurtado-Gil2017,kurban2023} respectively. Hence, we shall use these models to fit our data. Mathematically, these models can be represented as follows:
\begin{itemize}
    \item GH model:
    \[ \chi = \frac{1 }{1+\frac{1}{2}\overline{N} \xi} \]
    
    \item NB model:
    \[ \chi = \frac{\ln(1 + \overline{N} \xi)}{\overline{N} \xi} \]
\end{itemize}

\section{Simulations}\label{sim}

In this work, we use the publicly available halo and galaxy catalogues obtained from the \textsc{Quijote} N-body simulations \citep{Villaescusa-Navarro2020,Hahn2021}. The initial conditions were set using the codes \textsc{2LPTIC} \citep{Crocce2006} and \textsc{2LPTPNG} \citep{Scoccimarro2012, Coulton2023} at redshift $z = 127$, and the subsequent evolution of $512^{3}$ dark-matter particles was tracked to $z = 0$ using the code \textsc{Gadget-III} \citep{Springel2005}. In each simulation snapshot, the haloes were identified at $z=\{0,0.5,1,2,3\}$ using the friends-of-friends (FOF) algorithm \citep{Davis1985}, while the galaxy catalogs were constructed only at $z=0$ using a statistical Halo Occupation Distribution (HOD) prescription \citep{Zheng2007}, rather than being extracted from semi-analytic or full hydrodynamical simulations. All simulations have a cosmological volume of $1 (h^{-1}\text{Gpc})^{3}$.

We summarize the cosmological parameters of the simulations used for the construction of halo catalogs in Table~\ref{tab:halo} and for the galaxy catalogs in Table~\ref{tab:galaxy}, respectively. Each simulation consists of multiple realizations, typically ranging from $500$ to $15000$. For our analysis, we randomly select one realization from each simulation set. For simulations differing from the standard $\Lambda$CDM cosmology, a single parameter is slightly shifted from its fiducial value to assess its impact on the VPF. For the halo catalogues, we focus our analysis on the influence of the amplitudes of primordial non-Gaussianity, $\{f_{\mathrm{NL}}^{\mathrm{local}}, f_{\mathrm{NL}}^{\mathrm{equil}}, f_{\mathrm{NL}}^{\mathrm{ortho}}\}$, total neutrino mass $\{M_\nu\}$, the background density contrast parameter $\{\delta_{\rm b}\}$, the dark energy parameter $\{\omega\}$, and the $f(R)$ modified gravity parameter $\{f_{R_0}\}$ using the Hu \& Sawicki model \citep{Hu2007}. While for the galaxy catalogues, we focus on the five vanilla $\Lambda$CDM parameters $\{\sigma_8, \Omega_{\mathrm{m}}, \Omega_{\mathrm{b}}, n_s, h\}$, along with total neutrino mass $\{M_\nu\}$. For a more detailed description of these cosmological parameters, see \cite{Villaescusa-Navarro2020} and the \textsc{Quijote} documentation \footnote{\url{https://quijote-simulations.readthedocs.io/en/latest/}}. We assess the impact of cosmic variance using multiple $\Lambda$CDM realizations in Appendix~\ref{cv}, where we find that cosmic variance introduces a non-negligible scatter and should be considered when interpreting differences in the VPF across cosmological models and observations.

We calculate the VPF for the dark-matter halo and the galaxy catalogues using the prescription described in \cite{fry2013}. Firstly, we divide the simulation volume into a regular grid of non-overlapping spherical cells of radius $R$. The probability of obtaining no tracers, $P_0(R)$ within $R$ can be written as
\begin{equation}
P_0(R) = \frac{N_{\rm empty}}{N_{\rm total}},
\end{equation}
where $N_{\rm empty}$ is the number of cells containing zero tracers and $N_{\rm total}$ is the total number of cells.
The volume-averaged two-point correlation function is estimated from the excess variance relative to Poisson as follows \cite{fry2013},
\begin{equation}\label{sc.var}
\overline{N}^2 \, \overline{\xi}_2 = \langle N^2 \rangle - \overline{N}^2 - \overline{N},
\end{equation}
where $\overline{N} = \langle N \rangle$ is the mean number of tracers within $R$. We obtain the void statistic $\chi(R)$ as follows,
\begin{equation}
\chi(R) = -\frac{\ln P_0(R)}{\overline{N}},
\end{equation}
and is analyzed as a function of the scaling variable $\overline{N}\xi$ (Eq.~\ref{sc.var}) in the range \( R = 0.5\text{--}25\,h^{-1}\mathrm{Mpc} \) in steps of $0.5 \,h^{-1}\mathrm{Mpc}$ with the errors on $P_{0}$, and \( \chi \) estimated as follows \citep{hamilton1985,Maurogordato1987,Colombi1995},

\begin{equation}
    \Delta P_{0} = \sqrt{\frac{P_{0}(1-P_{0})}{N_{total}}},
\end{equation}
and
\begin{equation}
 \bigg( \frac{\Delta \chi}{\chi}\bigg)   = \bigg |\frac{\Delta P_{0}}{P_{0} | \log P_{0}|} - \frac{\Delta \overline{N} }{\overline{N}} \bigg |.
\end{equation}

\begin{table}[ht]
\centering
\caption{Cosmological parameters of \textsc{Quijote} simulations used for halo catalogues.}
\resizebox{\textwidth}{!}{%
\begin{tabular}{|l||c|c|c|c|c|c|c|c|c|c|c|c|}
\hline
\hline
Cosmology & $\Omega_{\rm m}$ & $\Omega_{\rm b}$ & $h$ & $n_{\rm s}$ & $\sigma_8$ & $M_\nu$ (eV) & $\omega$ & $\delta_{\rm b}$ & $f_{\rm NL}^{\rm local}$ & $f_{\rm NL}^{\rm equil.}$ & $f_{\rm NL}^{\rm ortho\text{-}LSS}$ & $f_{R_0}$ \\
\hline
\hline
$\Lambda$CDM & $0.3175$ & $0.049$ & $0.6711$ & $0.9624$ & $0.834$ & $0$ & $-1$ & $0$ & $0$ & $0$ & $0$ & $0$ \\
\hline
$\text{DC}\_\text{m}$ & $0.3175$ & $0.049$ & $0.6711$ & $0.9624$ & $0.834$ & $0$ & $-1$ & $-0.035$ & $0$ & $0$ & $0$ & $0$ \\
\hline
$\text{DC}\_\text{p}$ & $0.3175$ & $0.049$ & $0.6711$ & $0.9624$ & $0.834$ & $0$ & $-1$ & $+0.035$ & $0$ & $0$ & $0$ & $0$ \\
\hline
$\text{Eq}\_\text{m}$ & $0.3175$ & $0.049$ & $0.6711$ & $0.9624$ & $0.834$ & $0$ & $-1$ & $0$ & $0$ & $-100$ & $0$ & $0$ \\
\hline
$\text{Eq}\_\text{p}$ & $0.3175$ & $0.049$ & $0.6711$ & $0.9624$ & $0.834$ & $0$ & $-1$ & $0$ & $0$ & $+100$ & $0$ & $0$ \\
\hline
$\text{fR}\_\text{pppp}$ & $0.3175$ & $0.049$ & $0.6711$ & $0.9624$ & $0.834$ & $0$ & $-1$ & $0$ & $0$ & $0$ & $0$ & $-5e-4$ \\
\hline
$\text{Mnu}\_\text{ppp}$ & $0.3175$ & $0.049$ & $0.6711$ & $0.9624$ & $0.834$ & $0.4$ & $-1$ & $0$ & $0$ & $0$ & $0$ & $0$ \\
\hline
$\text{LC}\_\text{m}$ & $0.3175$ & $0.049$ & $0.6711$ & $0.9624$ & $0.834$ & $0$ & $-1$ & $0$ & $-100$ & $0$ & $0$ & $0$ \\
\hline
$\text{LC}\_\text{p}$ & $0.3175$ & $0.049$ & $0.6711$ & $0.9624$ & $0.834$ & $0$ & $-1$ & $0$ & $+100$ & $0$ & $0$ & $0$ \\
\hline
$\text{OR}\_\text{LSS}\_\text{m}$ & $0.3175$ & $0.049$ & $0.6711$ & $0.9624$ & $0.834$ & $0$ & $-1$ & $0$ & $0$ & $0$ & $-100$ & $0$ \\
\hline
$\text{OR}\_\text{LSS}\_\text{p}$ & $0.3175$ & $0.049$ & $0.6711$ & $0.9624$ & $0.834$ & $0$ & $-1$ & $0$ & $0$ & $0$ & $+100$ & $0$ \\
\hline
$\text{w}\_\text{m}$ & $0.3175$ & $0.049$ & $0.6711$ & $0.9624$ & $0.834$ & $0$ & $-0.95$ & $0$ & $0$ & $0$ & $0$ & $0$ \\
\hline
$\text{w}\_\text{p}$ & $0.3175$ & $0.049$ & $0.6711$ & $0.9624$ & $0.834$ & $0$ & $-1.05$ & $0$ & $0$ & $0$ & $0$ & $0$ \\
\hline
\end{tabular}
}
\label{tab:halo}
\end{table}

\begin{table}[ht]
\centering
\caption{Same as Table~\ref{tab:halo} but for galaxy catalogues.}
\resizebox{\textwidth}{!}{%
\begin{tabular}{|l||c|c|c|c|c|c|c|c|c|c|c|c|}
\hline
\hline
Cosmology & $\Omega_{\rm m}$ & $\Omega_{\rm b}$ & $h$ & $n_{\rm s}$ & $\sigma_8$ & $M_\nu$ (eV) & $\omega$ & $\delta_{\rm b}$ & $f_{\rm NL}^{\rm local}$ & $f_{\rm NL}^{\rm equil.}$ & $f_{\rm NL}^{\rm ortho\text{-}LSS}$ & $f_{R_0}$ \\
\hline
\hline
$\Lambda$CDM & $0.3175$ & $0.049$ & $0.6711$ & $0.9624$ & $0.834$ & $0$ & $-1$ & $0$ & $0$ & $0$ & $0$ & $0$ \\
\hline
$\text{Mnu}\_\text{ppp}$ & $0.3175$ & $0.049$ & $0.6711$ & $0.9624$ & $0.834$ & $0.4$ & $-1$ & $0$ & $0$ & $0$ & $0$ & $0$ \\
\hline
$\text{h}\_\text{m}$ & $0.3175$ & $0.049$ & $0.6511$ & $0.9624$ & $0.834$ & $0$ & $-1$ & $0$ & $0$ & $0$ & $0$ & $0$ \\
\hline
$\text{h}\_\text{p}$ & $0.3175$ & $0.049$ & $0.6911$ & $0.9624$ & $0.834$ & $0$ & $-1$ & $0$ & $0$ & $0$ & $0$ & $0$ \\
\hline
$\text{ns}\_\text{m}$ & $0.3175$ & $0.049$ & $0.6711$ & $0.9424$ & $0.834$ & $0$ & $-1$ & $0$ & $0$ & $0$ & $0$ & $0$ \\
\hline
$\text{ns}\_\text{p}$ & $0.3175$ & $0.049$ & $0.6711$ & $0.9824$ & $0.834$ & $0$ & $-1$ & $0$ & $0$ & $0$ & $0$ & $0$ \\
\hline
$\text{Ob2}\_\text{m}$ & $0.3175$ & $0.047$ & $0.6711$ & $0.9624$ & $0.834$ & $0$ & $-1$ & $0$ & $0$ & $0$ & $0$ & $0$ \\
\hline
$\text{Ob2}\_\text{p}$ & $0.3175$ & $0.051$ & $0.6711$ & $0.9624$ & $0.834$ & $0$ & $-1$ & $0$ & $0$ & $0$ & $0$ & $0$ \\
\hline
$\text{Om}\_\text{m}$ & $0.3075$ & $0.049$ & $0.6711$ & $0.9624$ & $0.834$ & $0$ & $-1$ & $0$ & $0$ & $0$ & $0$ & $0$ \\
\hline
$\text{Om}\_\text{p}$ & $0.3275$ & $0.049$ & $0.6711$ & $0.9624$ & $0.834$ & $0$ & $-1$ & $0$ & $0$ & $0$ & $0$ & $0$ \\
\hline
$\text{s8}\_\text{m}$ & $0.3175$ & $0.049$ & $0.6711$ & $0.9624$ & $0.819$ & $0$ & $-1$ & $0$ & $0$ & $0$ & $0$ & $0$ \\
\hline
$\text{s8}\_\text{p}$ & $0.3175$ & $0.049$ & $0.6711$ & $0.9624$ & $0.849$ & $0$ & $-1$ & $0$ & $0$ & $0$ & $0$ & $0$ \\
\hline
\end{tabular}
}

\label{tab:galaxy}
\end{table}

\section{VPF in Dark Matter Haloes}\label{vpf:dm}
\subsection{VPF dependence on redshift}

\begin{figure}[htbp]
\centering
\includegraphics[width=.325\textwidth]{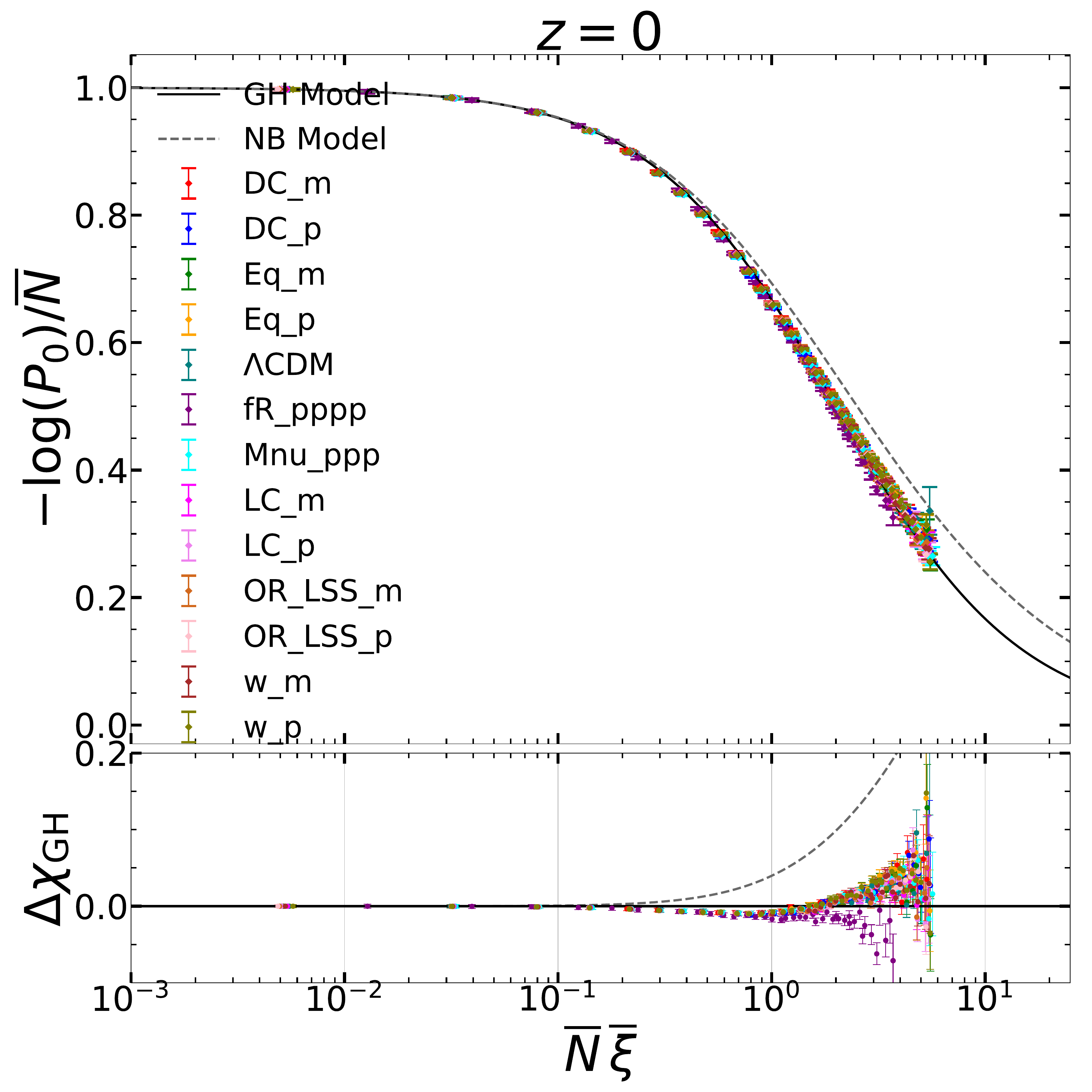}
\includegraphics[width=.325\textwidth]{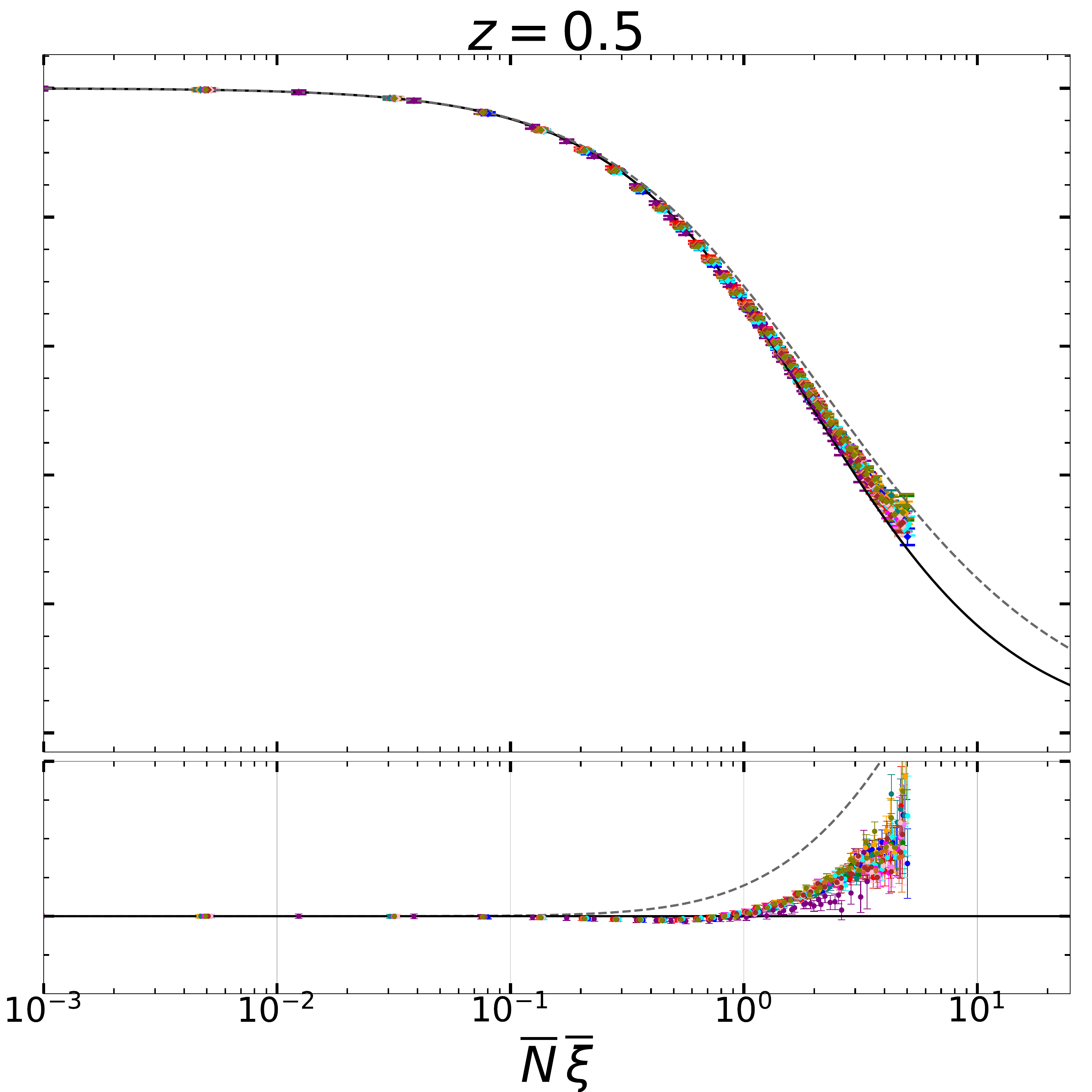}
\includegraphics[width=.325\textwidth]{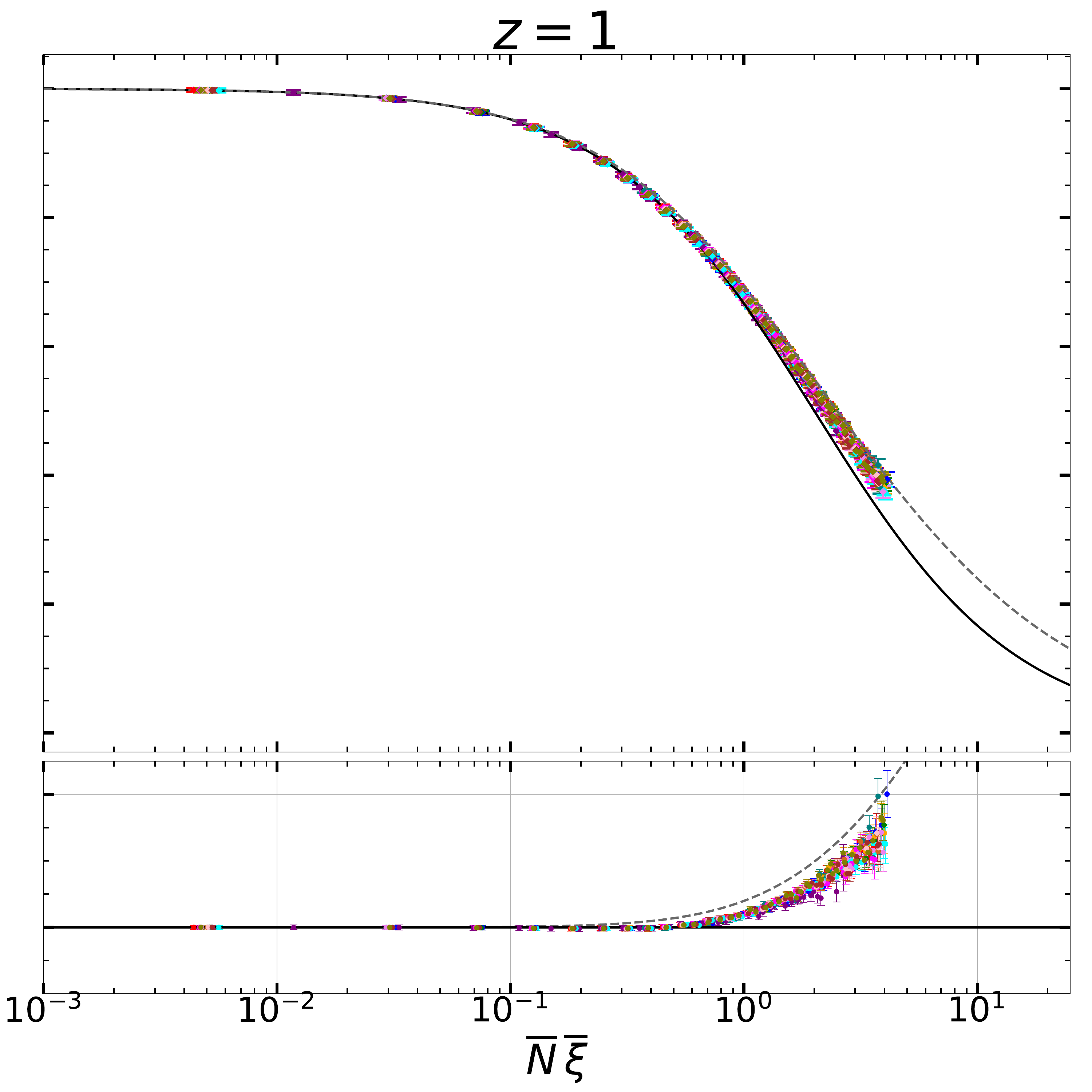}
\includegraphics[width=.325\textwidth]{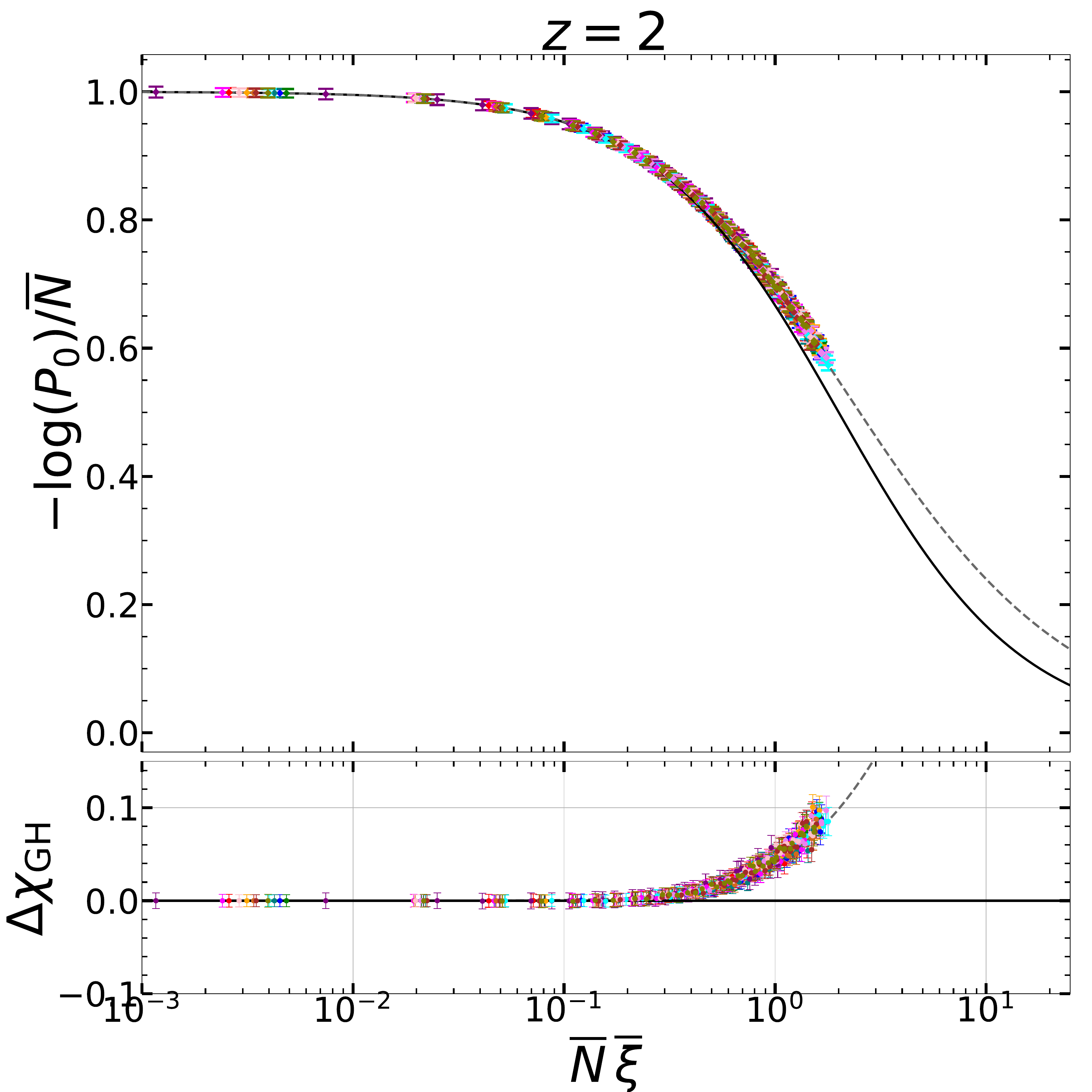}
\includegraphics[width=.325\textwidth]{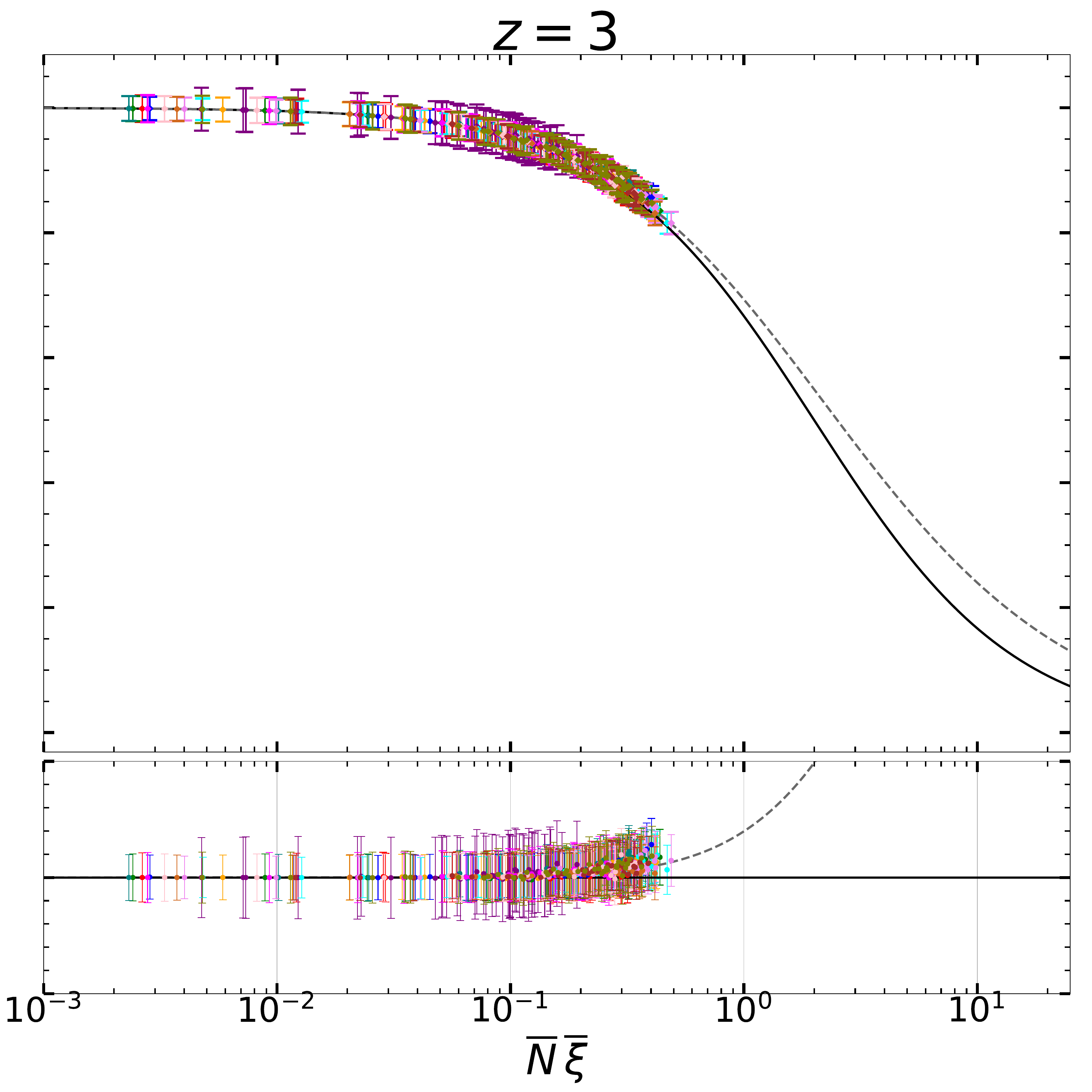}
\caption{VPF of the dark matter haloes at $z=0$ (upper left), $z=0.5$ (upper middle), $z=1$ (upper right), $z=2$ (lower left), $z=3$ (lower right). The solid line represents the analytical estimation of $\chi = -\log P_{0}/\overline{N}$ with the GH model, while the dashed line represents the same but for the NB model. The colored markers represent the $\chi$ obtained for the various dark matter catalogues listed in Table~\ref{tab:halo}. The bottom panel in each figure shows the difference between the $\chi$ obtained in each dark matter halo catalogue and that of the GH model.   \label{halo:z}}
\end{figure}

Figure~\ref{halo:z} presents the VPF of dark matter haloes at various $z$. At $z = 0$, the VPF closely follows the predictions of the GH model. The lower panels in each subfigure illustrate the difference between the $\chi$ obtained in each dark-matter halo catalogue and that of the GH model, with the calculation of uncertainties described in Appendix~\ref{un}. These panels demonstrate that, with increasing $z$, the VPF progressively deviates from the GH model and tends to follow the NB model. Furthermore, Table~\ref{tab:z_halo} highlights that, at $z = 0$, the deviation in the $\Lambda$CDM cosmology has the lowest statistical significance, at $\sim2\sigma$. In contrast, the $\text{Mnu}\_\text{ppp}$ model, which incorporates massive neutrinos with a total mass of $0.4$ eV, shows the most significant deviation, at $\sim7\sigma$. The strong deviation of the $\text{Mnu}\_\text{ppp}$ model may be attributed to the presence of neutrino particles with high thermal velocities and a scale-dependent growth of structures which might suppress small-scale clustering and produce a smoother matter distribution \citep{Lesgourgues2006,Navarro2015,Castorina2015}, and could therefore produce a void distribution that strongly deviates from the GH model, which is based on a scale-invariant, strongly clustered halo distribution.

We further observe that up to $z = 2$, the departure from the GH model becomes increasingly significant across all cosmologies, with increasing statistical significance in the $\chi$ residuals. However, at $z = 3$, the deviation from the GH model is unusually reduced, mainly due to the drastic decline in the number density of haloes by more than two orders of magnitude compared to $z=0$. This limits the analysis to only small values of $\overline{N}\xi$, which tends toward the Poisson limit, and thus reduces the apparent deviations from the theoretical models.

\begin{table}[ht]
\centering
\caption{Redshift evolution of VPF in dark matter haloes.}
\resizebox{\textwidth}{!}
{
\begin{tabular}{|l||c|c|c|c|c|}
\hline
\hline
\multicolumn{1}{|c||}{} & \multicolumn{5}{c|}{Deviations from the GH model at different $z$} \\
\cline{2-6}
\multicolumn{1}{|c||}{Cosmology} & $z=0$ & $z=0.5$ & $z=1$ & $z=2$ & $z=3$  \\
\hline
\hline
$\Lambda$CDM & $0.0689 \pm 0.0340$ & $0.0686 \pm 0.0088$ & $0.0804 \pm 0.0055$ & $0.0440 \pm 0.0026$ & $0.0108 \pm 0.0023$ \\
\hline
$\text{DC}\_\text{m}$ & $0.0239 \pm 0.0054$ & $0.0551 \pm 0.0083$  & $0.0718 \pm 0.0050$ & $0.0436 \pm 0.0026$ & $0.0060 \pm 0.0018$ \\
\hline
$\text{DC}\_\text{p}$ & $0.0307 \pm 0.0061$ & $0.059 \pm 0.0062$  & $0.0889 \pm 0.0071$ & $0.0470 \pm 0.0027$ & $0.0105 \pm 0.0023$ \\
\hline
$\text{Eq}\_\text{m}$ & $0.0345 \pm 0.0084$ & $0.0732 \pm 0.0133$  & $0.0825 \pm 0.0057$ & $0.0496 \pm 0.0028$ & $0.0090 \pm 0.0022$ \\
\hline
$\text{Eq}\_\text{p}$ & $0.0381 \pm 0.0091$ & $0.0764 \pm 0.0113$ & $0.0831 \pm 0.0055$ & $0.0517 \pm 0.0028$ & $0.0089 \pm 0.0021$ \\
\hline
$\text{fR}\_\text{pppp}$ & $0.0268 \pm 0.0041$ & $0.0268 \pm 0.0041$ & $0.0440 \pm 0.0034$ & $0.0277 \pm 0.0021$ & $0.0047 \pm 0.0025$ \\
\hline
$\text{Mnu}\_\text{ppp}$ & $0.0219 \pm 0.0031$  & $0.0574 \pm 0.0078$ & $0.0742 \pm 0.0050$ & $0.0513 \pm 0.0027$ & $0.0091 \pm 0.0019$ \\
\hline
$\text{LC}\_\text{m}$ & $0.0197 \pm 0.0063$ & $0.0510 \pm 0.0070$ & $0.0660 \pm 0.0045$ & $0.0390 \pm 0.0024$ & $0.0067 \pm 0.0020$ \\
\hline
$\text{LC}\_\text{p}$ & $0.0222 \pm 0.0035$ & $0.0538 \pm 0.0076$ & $0.0752 \pm 0.0049$ & $0.0523 \pm 0.0028$ & $0.0095 \pm 0.0020$ \\
\hline
$\text{OR}\_\text{LSS}\_\text{m}$ & $0.0222 \pm 0.0062$ & $0.0502 \pm 0.0057$ & $0.0729 \pm 0.0049$ & $0.0441 \pm 0.0026$ & $0.0062 \pm 0.0017$ \\
\hline
$\text{OR}\_\text{LSS}\_\text{p}$ & $0.0240 \pm 0.0046$ & $0.0538 \pm 0.0068$ & $0.0734 \pm 0.0050$ & $0.0469 \pm 0.0026$ & $0.0076 \pm 0.0020$ \\
\hline
$\text{w}\_\text{m}$ & $0.0207 \pm 0.0043$ & $0.0513 \pm 0.0069$ & $0.0727 \pm 0.0050$ & $0.0451 \pm 0.0026$ & $0.0073 \pm 0.0019$ \\
\hline
$\text{w}\_\text{p}$ & $0.0380 \pm 0.0096$ & $0.0844 \pm 0.0138$ & $0.0839 \pm 0.0057$ & $0.0458 \pm 0.0026$ & $0.0090 \pm 0.0023$ \\
\hline
\end{tabular}
}
\label{tab:z_halo}
\end{table}

\subsection{VPF dependence on mass}

\begin{figure}[htbp]
\centering
\includegraphics[width=.325\textwidth]{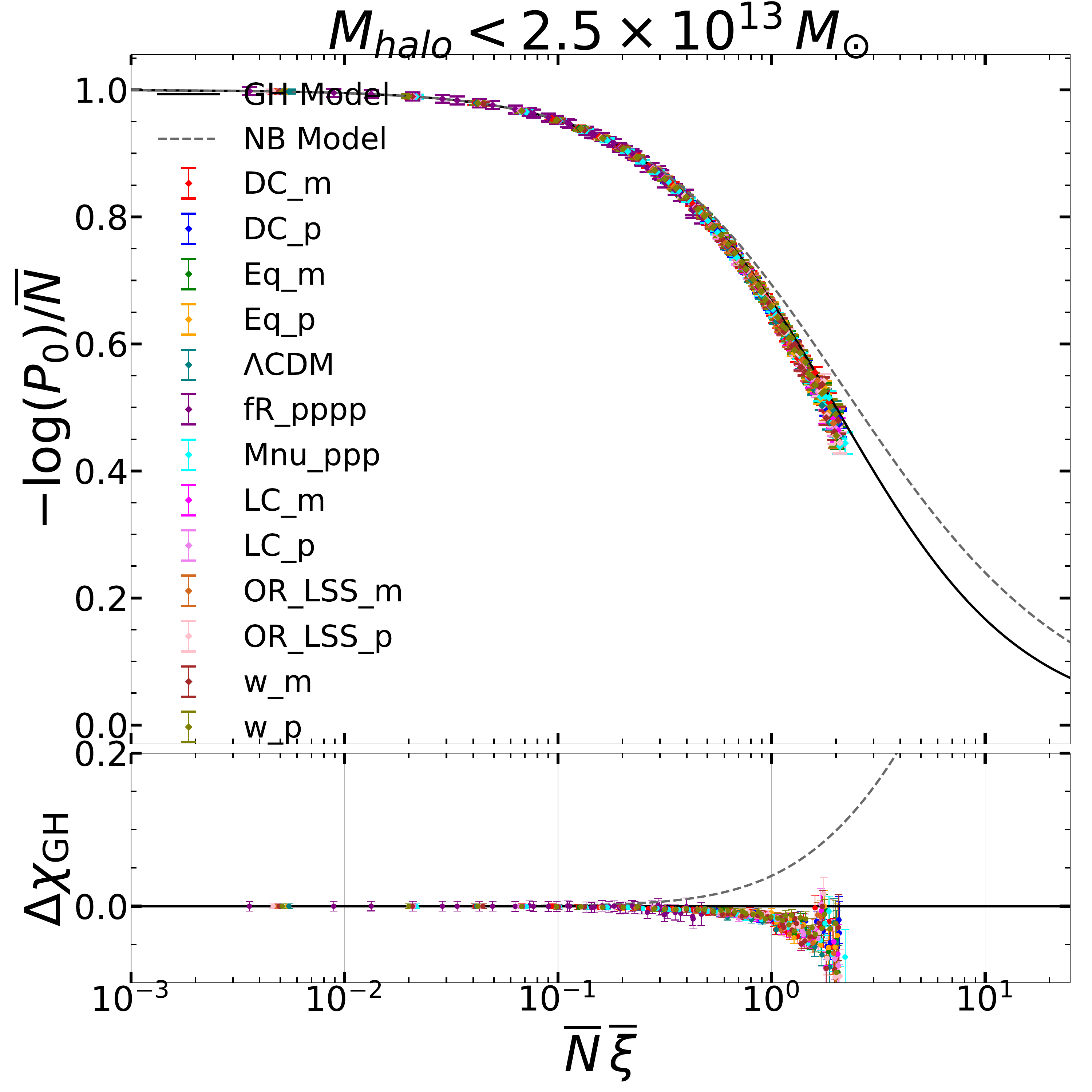}
\includegraphics[width=.325\textwidth]{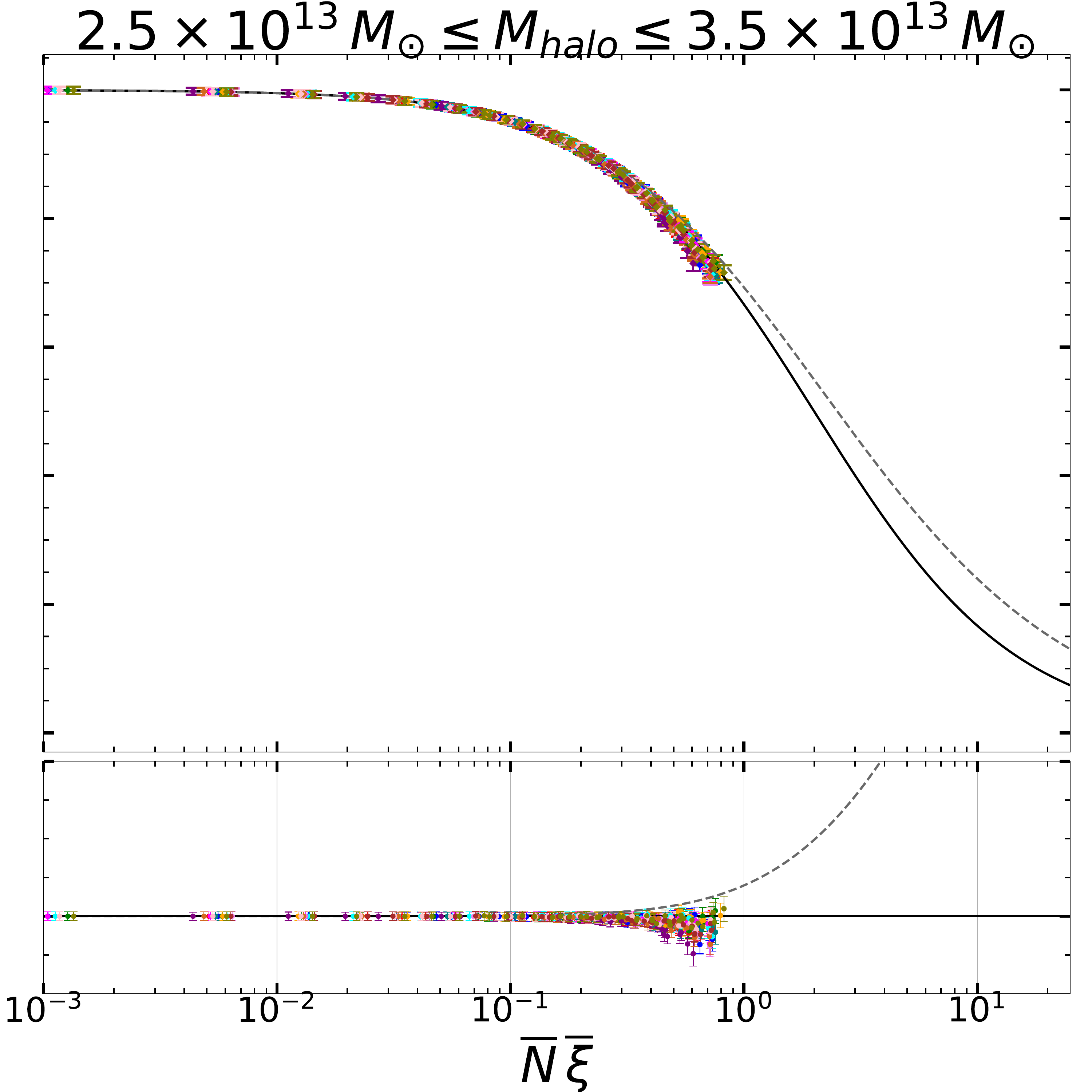}
\includegraphics[width=.325\textwidth]{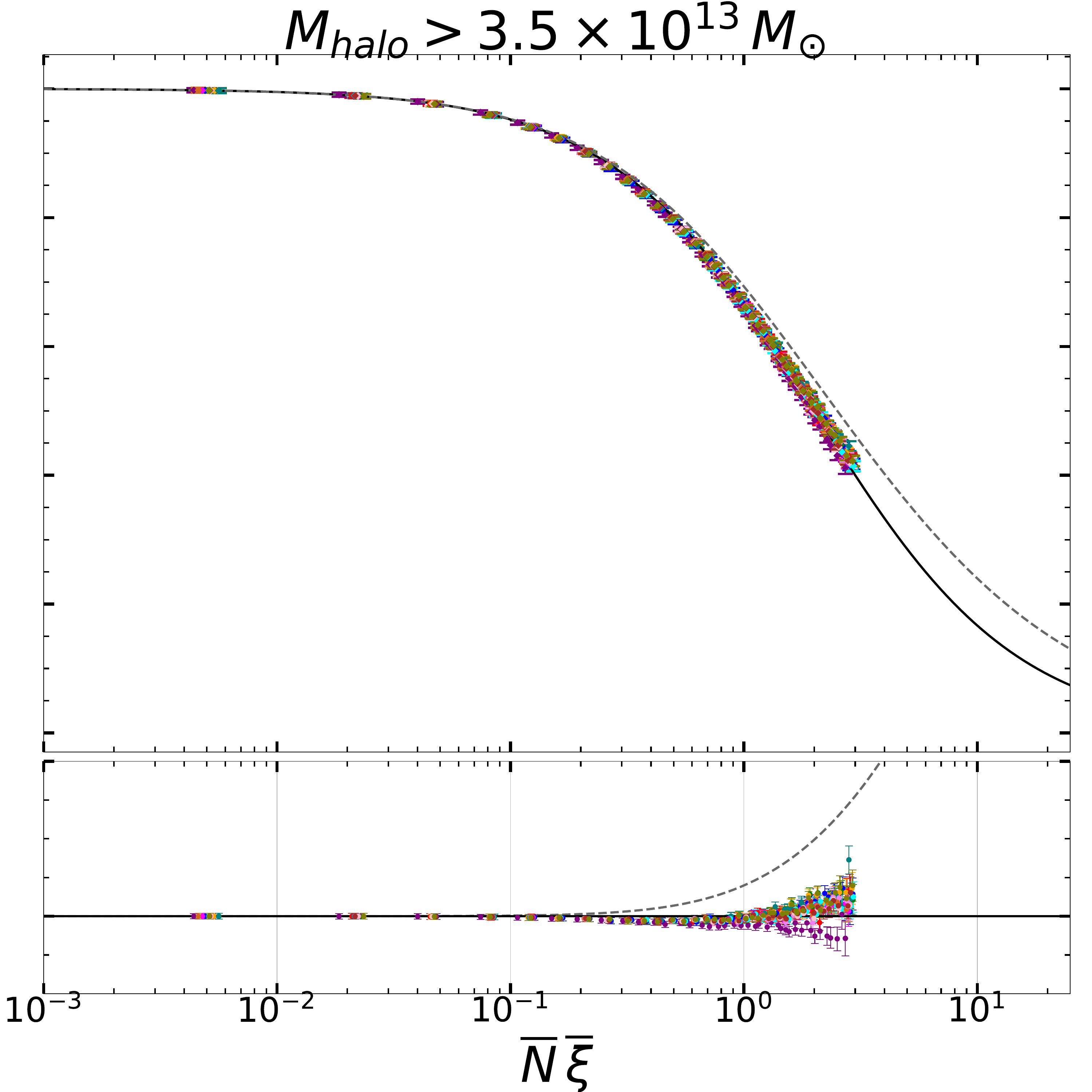}
\caption{Same as Figure~\ref{halo:z} but for the dark matter haloes at $z=0$ with masses $M_\mathrm{halo} < 2.5\times 10^{13}~M_\odot$ (left), $2.5\times 10^{13}~M_\odot < M_\mathrm{halo} < 3.5\times 10^{13}~M_\odot$ (middle), and  $M_\mathrm{halo} > 3.5\times 10^{13}~M_\odot$ (right). \label{halo:mass}}
\end{figure}

In Figure~\ref{halo:mass}, we show the dependence of the VPF on the mass of dark matter haloes ($M_\mathrm{halo}$) in three mass bins: low mass ($M_\mathrm{halo} < 2.5 \times 10^{13}~M_\odot$),
intermediate mass ($2.5 \times 10^{13}~M_\odot < M_\mathrm{halo} < 3.5 \times 10^{13}~M_\odot$),
and high mass ($M_\mathrm{halo} > 3.5 \times 10^{13}~M_\odot$). The minimum halo mass in the catalogue is $M_\mathrm{halo} \sim 1.3 \times 10^{13}~M_\odot$, set by the resolution of the \textsc{Quijote} simulations. These mass thresholds were chosen to ensure a sufficient number of halos in each bin, enabling robust estimation of the VPF in each case.

We observe that in all three mass bins, the VPF in all cosmological models tends to follow the GH model. From Table~\ref{tab:halo_mass}, which summarizes the deviations of the VPF from the GH model, we find that the deviation varies systematically with both halo mass and cosmology. In particular, in the low-mass bin, the standard $\Lambda$CDM and the cosmology with $\omega = -1.05$ $(\text{w}\_\text{p})$ show some of the most statistically significant deviations. On the other hand, the cosmology with modified gravity ($\text{fR}\_\text{pppp}$) shows the smallest deviations, indicating that the low-mass haloes in this modified gravity scenario are more consistent with the predictions of the GH model. For the intermediate-mass bin, we see a sharp increase in deviation across all cosmologies, with $\text{fR}\_\text{pppp}$ showing a particularly significant deviation, which suggests that the intermediate-mass haloes are more sensitive to the effects of modified gravity, leading to stronger deviations from standard hierarchical clustering, consistent with previous works \citep{Arnalte-Mur2017}. For the high-mass bins, the $\text{fR}\_\text{pppp}$ cosmology continues to show a relatively large deviation, likely due to the altered gravitational dynamics in dense environments where massive haloes reside. Meanwhile, cosmologies like $\text{LC}\_\text{m}$ and $\text{OR}\_\text{LSS}\_\text{m}$ show much smaller deviations, which indicate that the VPF for the high-mass haloes is less sensitive to the effects of the amplitudes of primordial non-Gaussianity. This behaviour is consistent with previous studies of halo clustering in the presence of primordial non-Gaussianity, which find a clear mass-dependent response, with reduced sensitivity at the high-mass end \citep{guti2024}.

\begin{table}[ht]
\centering
\caption{VPF dependence on the halo mass.}
\resizebox{\textwidth}{!}
{
\begin{tabular}{|l||c|c|c|}
\hline
\hline
\multicolumn{1}{|c||}{} & \multicolumn{3}{c|}{Deviations from the GH model at different mass ranges} \\
\cline{2-4}
\multicolumn{1}{|c||}{Cosmology} & $M_\mathrm{halo} < 2.5\times 10^{13}~M_\odot$ & $2.5\times 10^{13}~M_\odot < M_\mathrm{halo} < 3.5\times 10^{13}~M_\odot$ & $M_\mathrm{halo} > 3.5\times 10^{13}~M_\odot$ \\
\hline
\hline
$\Lambda$CDM & $0.0295 \pm 0.0037$ & $0.0091 \pm 0.0012$ & $0.0184 \pm 0.0022$ \\
\hline
$\text{DC}\_\text{m}$ & $0.0232 \pm 0.0035$ & $0.0092 \pm 0.0011$ & $0.0128 \pm 0.0019$ \\
\hline
$\text{DC}\_\text{p}$ & $0.0225 \pm 0.0033$ & $0.0119 \pm 0.0014$ & $0.0159 \pm 0.0018$ \\
\hline
$\text{Eq}\_\text{m}$ & $0.0315 \pm 0.0048$ & $0.0064 \pm 0.0008$ & $0.0144 \pm 0.0016$ \\
\hline
$\text{Eq}\_\text{p}$ & $0.0256 \pm 0.0036$ & $0.0051 \pm 0.0007$ & $0.0162 \pm 0.0020$ \\
\hline
$\text{fR}\_\text{pppp}$ & $0.0061 \pm 0.0009$ & $0.0176 \pm 0.0018$ & $0.0157 \pm 0.0017$ \\
\hline
$\text{Mnu}\_\text{ppp}$ & $0.0348 \pm 0.0045$ & $0.0091 \pm 0.0013$ & $0.0100 \pm 0.0015$ \\
\hline
$\text{LC}\_\text{m}$ & $0.0307 \pm 0.0043$ & $0.0086 \pm 0.0012$ & $0.0079 \pm 0.0011$ \\
\hline
$\text{LC}\_\text{p}$ & $0.0333 \pm 0.0047$ & $0.0144 \pm 0.0016$ & $0.0103 \pm 0.0014$ \\
\hline
$\text{OR}\_\text{LSS}\_\text{m}$ & $0.0348 \pm 0.0051$ & $0.0139 \pm 0.0015$ & $0.0082 \pm 0.0012$ \\
\hline
$\text{OR}\_\text{LSS}\_\text{p}$ & $0.0370 \pm 0.0051$ & $0.0078 \pm 0.0010$ & $0.0083 \pm 0.0012$ \\
\hline
$\text{w}\_\text{m}$ & $0.0304 \pm 0.0038$ & $0.0100 \pm 0.0012$ & $0.0107 \pm 0.0017$ \\
\hline
$\text{w}\_\text{p}$ & $0.0269 \pm 0.0048$ & $0.0059 \pm 0.0009$ & $0.0172 \pm 0.0020$ \\
\hline
\end{tabular}
}
\label{tab:halo_mass}
\end{table}

\subsubsection{VPF dependence on halo subsampling}

\begin{figure}[htbp]
\centering
\includegraphics[width=.325\textwidth]{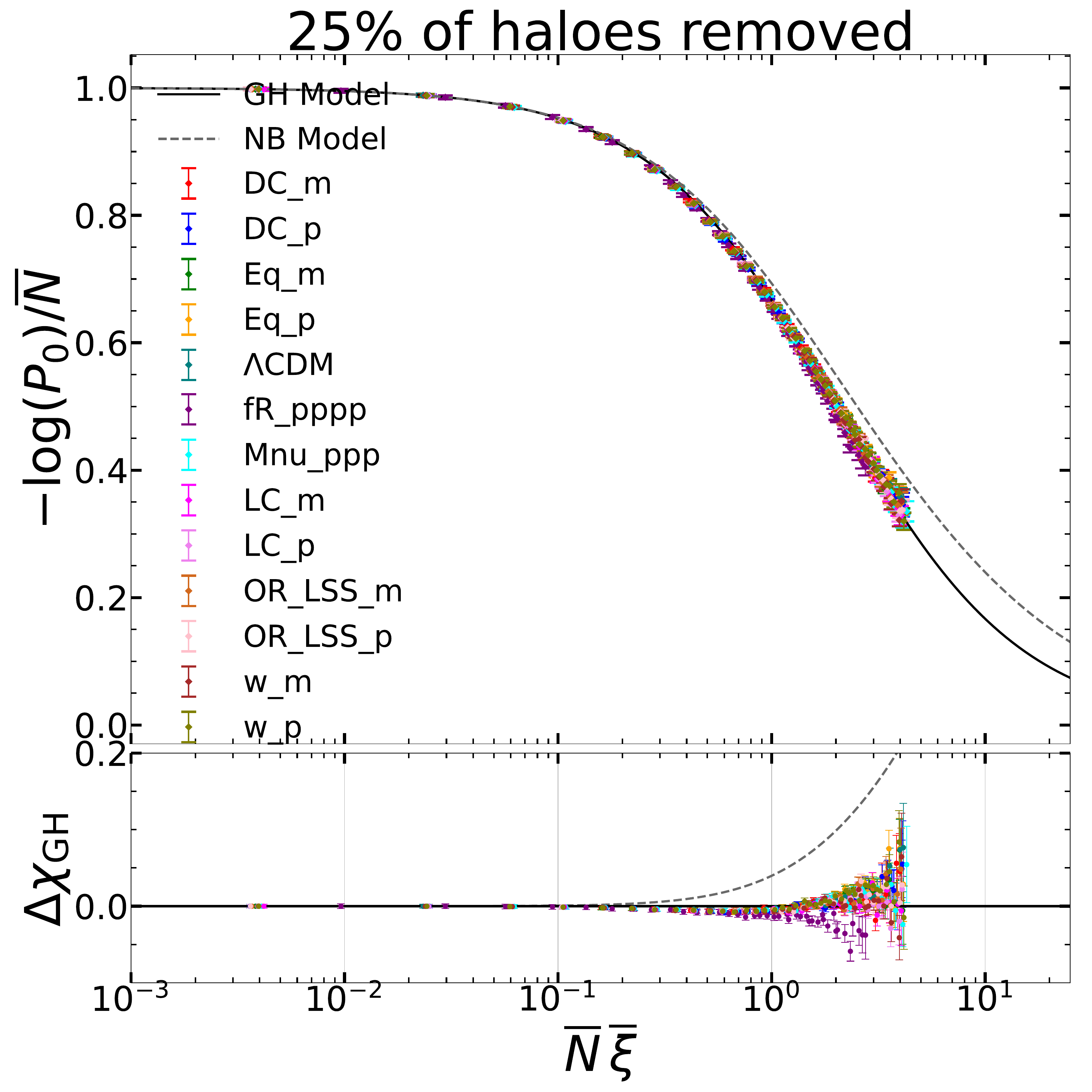}
\includegraphics[width=.325\textwidth]{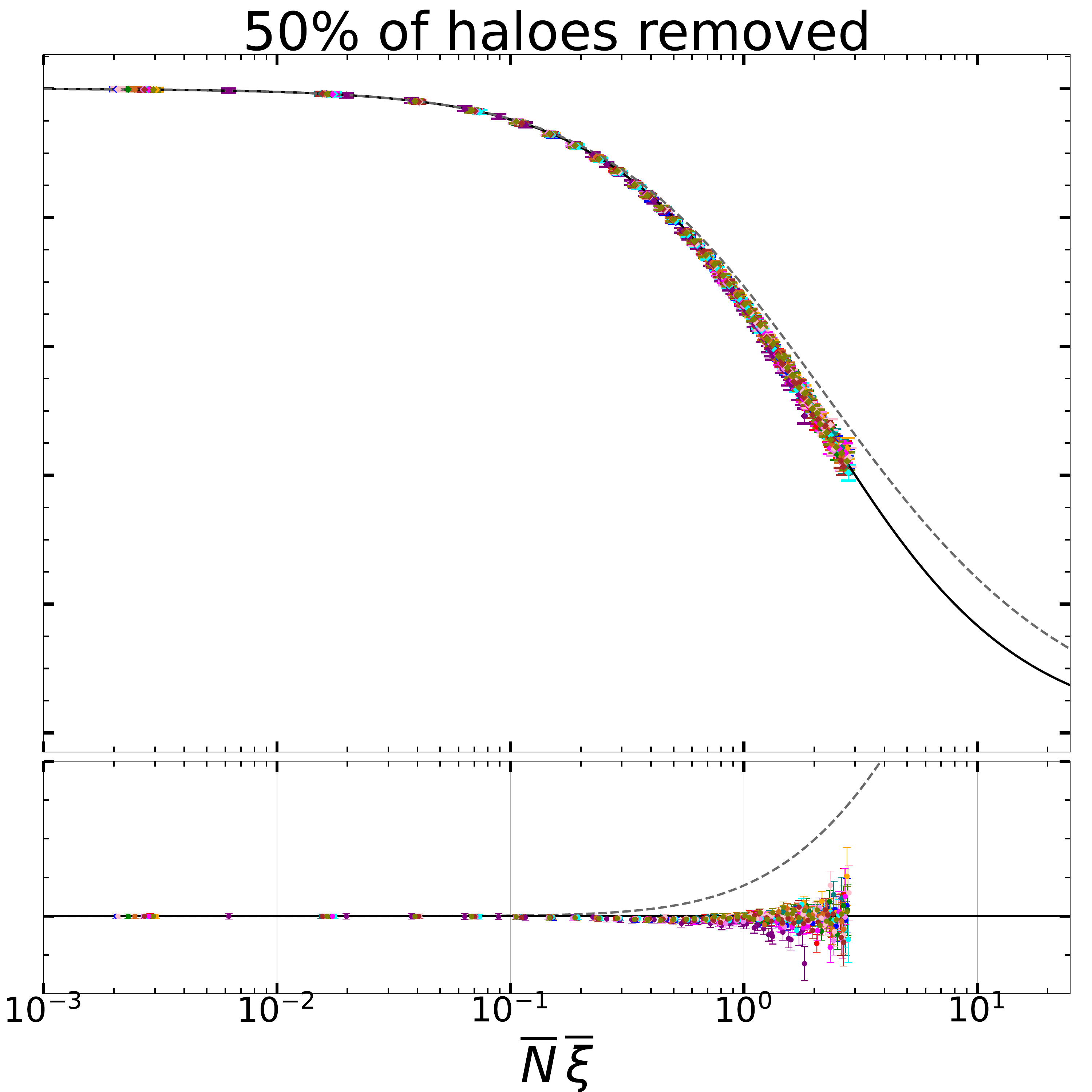}
\includegraphics[width=.325\textwidth]{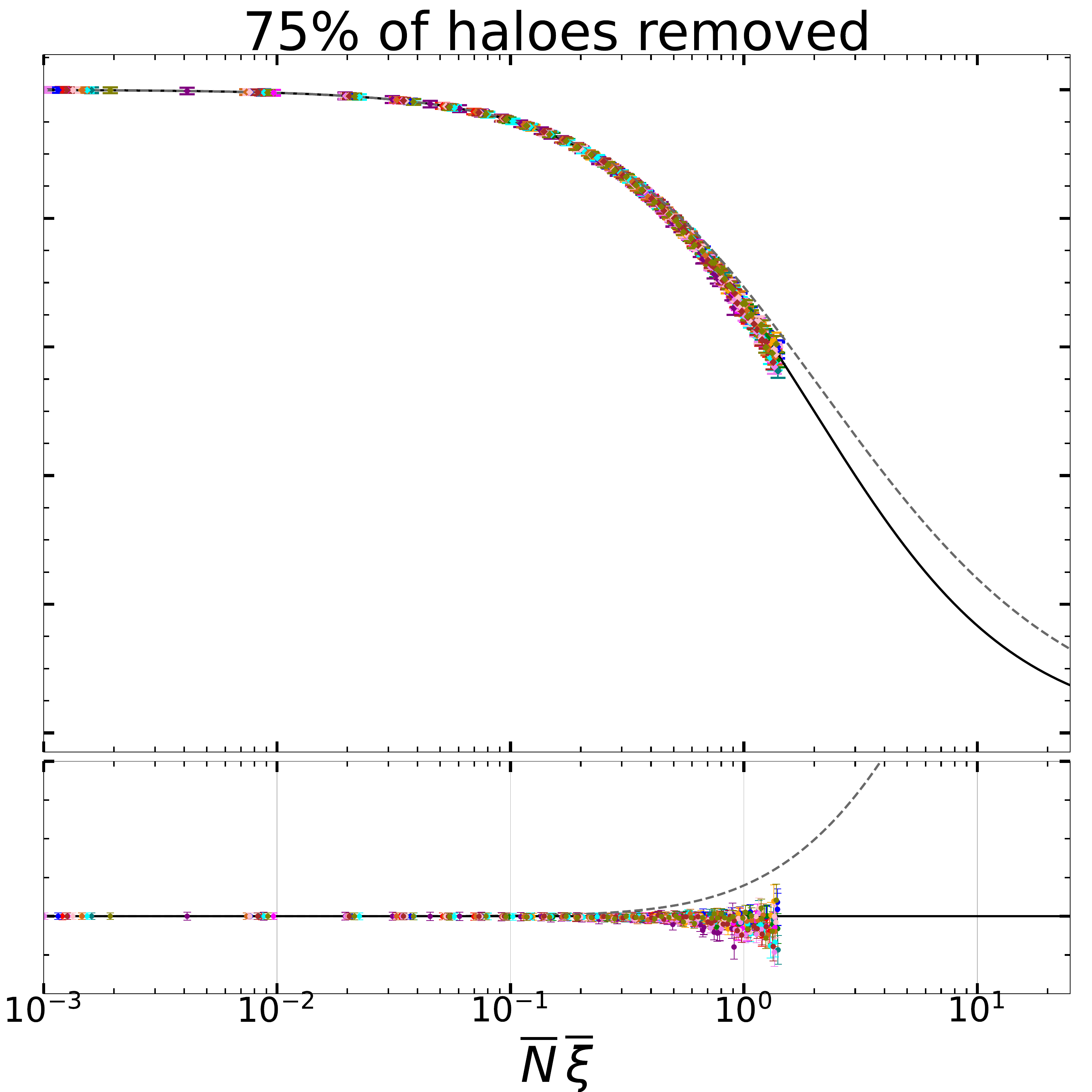}
\caption{Same as Figure~\ref{halo:z} but for the dark matter haloes at $z=0$ with random subsampling by $25\%$ (left), $50\%$ (center) and $75\%$ (right). \label{halo:sub}}
\end{figure}

Following \cite{croton2004b,kurban2023}, Figure~\ref{halo:sub} illustrates the impact of halo number density on the VPF by random subsampling at three levels: $25\%$, $50\%$, and $75\%$. We find that across all models and dilution levels, the VPF remains broadly consistent with the predictions of the GH model, though the degree of deviation increases with higher dilution. From Table~\ref{tab:halo:sub}, it is evident that the cosmology with modified gravity ($\text{fR}\_\text{pppp}$) exhibits relatively large deviations from the GH model at all levels of dilution. Other cosmologies, such as those with massive neutrinos ($\text{Mnu}\_\text{ppp}$), or with primordial non-Gaussianity ($\text{LC}_\text{m}$ and $\text{LC}_\text{p}$), also show increasing deviations with higher levels of subsampling.
This suggests that while the VPF of dark-matter haloes is generally robust under moderate reductions in tracer densities, higher levels of subsampling produce substantial deviations from the GH model in cosmologies with modified gravity, massive neutrinos, or primordial non-Gaussianity, highlighting the differences in clustering induced by these models more clearly.

\begin{table}[ht]
\centering
\caption{VPF dependence on the halo subsampling.}
\resizebox{0.8\textwidth}{!}
{
\begin{tabular}{|l||c|c|c|}
\hline
\hline
\multicolumn{1}{|c||}{} & \multicolumn{3}{c|}{Deviations from the GH model for different halo subsamples}  \\
\cline{2-4}
\multicolumn{1}{|c||}{Cosmology} & $25\%$ haloes removed & $50\%$ haloes removed & $75\%$ haloes removed \\
\hline
\hline
$\Lambda$CDM & $0.0252 \pm 0.0056$ & $0.0147 \pm 0.0030$ & $0.0073 \pm 0.0013$ \\
\hline
$\text{DC}\_\text{m}$ & $0.0217 \pm 0.0060$ & $0.0102 \pm 0.0021$ & $0.0076 \pm 0.0013$ \\
\hline
$\text{DC}\_\text{p}$ & $0.0172 \pm 0.0037$ & $0.0268 \pm 0.0070$ & $0.0110 \pm 0.0017$ \\
\hline
$\text{Eq}\_\text{m}$ & $0.0196 \pm 0.0036$ & $0.0068 \pm 0.0022$ & $0.0059 \pm 0.0012$ \\
\hline
$\text{Eq}\_\text{p}$ & $0.0225 \pm 0.0046$ & $0.0084 \pm 0.0022$ & $0.0059 \pm 0.0009$ \\
\hline
$\text{fR}\_\text{pppp}$ & $0.0277 \pm 0.0037$ & $0.0165 \pm 0.0022$ & $0.0112 \pm 0.0014$ \\
\hline
$\text{Mnu}\_\text{ppp}$ & $0.0128 \pm 0.0034$ & $0.0078 \pm 0.0011$ & $0.0119 \pm 0.0015$ \\
\hline
$\text{LC}\_\text{m}$ & $0.0137 \pm 0.0025$ & $0.0115 \pm 0.0026$ & $0.0078 \pm 0.0011$ \\
\hline
$\text{LC}\_\text{p}$ & $0.0156 \pm 0.0039$ & $0.0106 \pm 0.0025$ & $0.0086 \pm 0.0014$ \\
\hline
$\text{OR}\_\text{LSS}\_\text{m}$ & $0.0132 \pm 0.0026$ & $0.0081 \pm 0.0014$ & $0.0082 \pm 0.0014$ \\
\hline
$\text{OR}\_\text{LSS}\_\text{p}$ & $0.0157 \pm 0.0036$ & $0.0122 \pm 0.0021$ & $0.0082 \pm 0.0013$ \\
\hline
$\text{w}\_\text{m}$ & $0.0147 \pm 0.0029$ & $0.0070 \pm 0.0014$ & $0.0133 \pm 0.0018$ \\
\hline
$\text{w}\_\text{p}$ & $0.0174 \pm 0.0042$ & $0.0118 \pm 0.0025$ & $0.0085 \pm 0.0013$ \\
\hline
\end{tabular}
}
\label{tab:halo:sub}
\end{table}

\subsection{VPF dependence on redshift space}

\begin{figure}[htbp]
\centering
\includegraphics[width=.49\textwidth]{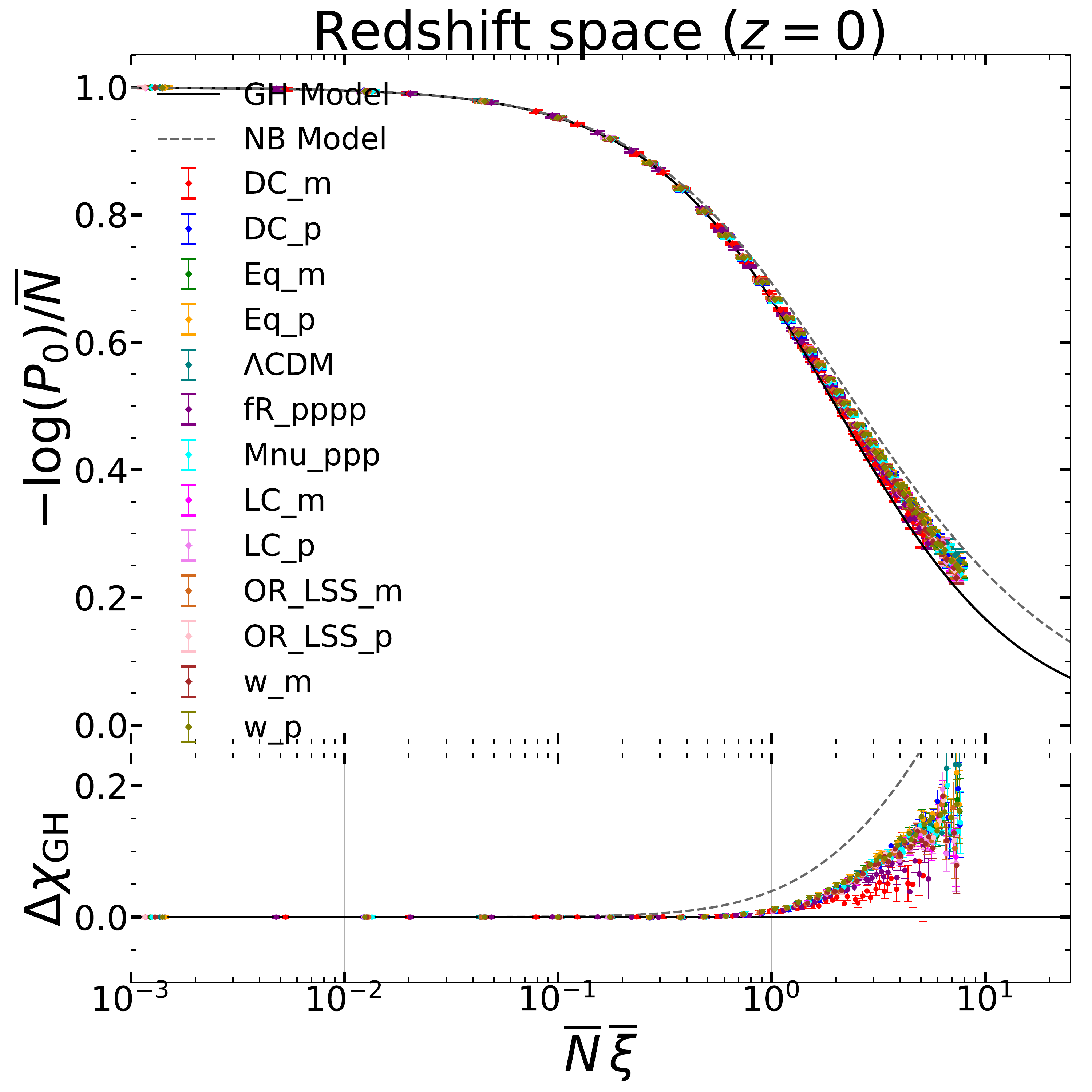}
\includegraphics[width=.49\textwidth]{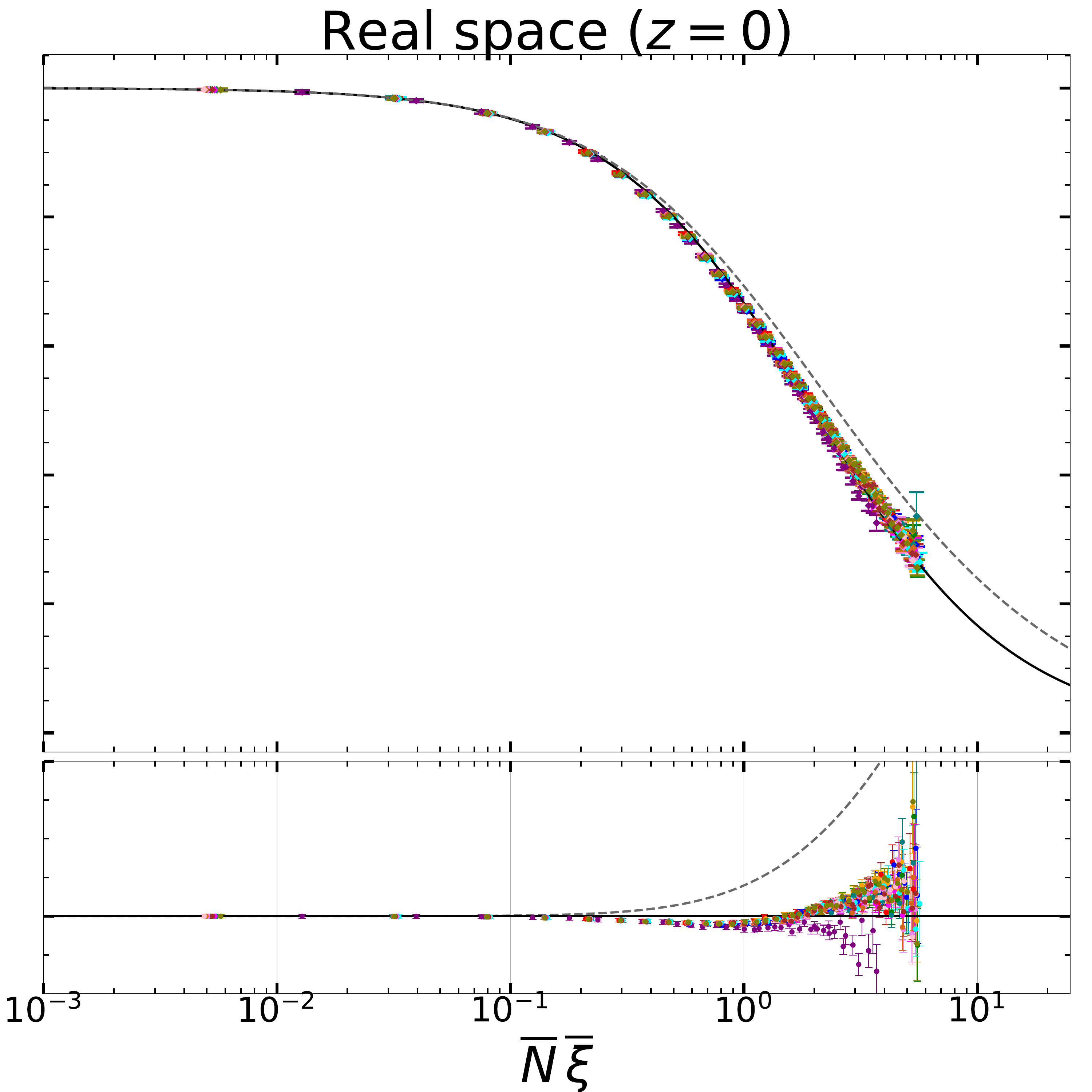}
\caption{Same as Figure~\ref{halo:z} but for the dark matter haloes in redshift space at $z=0$ (left), and real space (right). \label{halo:red}}
\end{figure}

In Figure~\ref{halo:red}, we compare the VPF measured in both real and redshift space. In real space, the VPF for all models follows the predictions of the GH model, while, in redshift space, the VPF deviates from the GH model at scales $\overline{N}\xi \gtrsim 1$ toward the NB model. This transition could be driven by the influence of peculiar velocities, which were previously suggested by \cite{croton2004} as a key factor causing the VPF to follow the NB model in redshift space. From Tables~\ref{tab:red} and \ref{tab:z_halo}, we observe that the cosmology with primordial non-Gaussianity ($\text{LC}\_\text{p}$) exhibits some of the largest deviations from the GH model in both real and redshift space. This suggests that the effects of redshift-space distortions are more pronounced in cosmological scenarios that incorporate primordial non-Gaussianities, and could be linked to their amplitudes. This behaviour is consistent with previous studies showing that primordial non-Gaussianity can modify the clustering of dark matter halos and velocity fields, thereby enhancing redshift-space distortions relative to the Gaussian case \citep{dalal2008}. In contrast, the $\Lambda$CDM cosmology shows relatively smaller deviations in both spaces, indicating a closer agreement with the GH predictions than the non-standard cosmologies considered here.

\begin{table}[ht]
\centering
\caption{Deviation of different cosmological models from the GH model in redshift space at $z=0$.}
\resizebox{0.7\textwidth}{!}
{
\begin{tabular}{|l||c|}
\hline
\hline
Cosmology & Deviations from the GH model in redshift space at $z=0$ \\
\hline
\hline
$\Lambda$CDM & $0.1158 \pm 0.0131$ \\
\hline
$\text{DC}\_\text{m}$ & $0.0389 \pm 0.0078$ \\
\hline
$\text{DC}\_\text{p}$ & $0.1032 \pm 0.0084$ \\
\hline
$\text{Eq}\_\text{m}$ & $0.1047 \pm 0.0087$ \\
\hline
$\text{Eq}\_\text{p}$ & $0.1121 \pm 0.0098$ \\
\hline
$\text{fR}\_\text{pppp}$ & $0.0514 \pm 0.0041$ \\
\hline
$\text{Mnu}\_\text{ppp}$ & $0.0990 \pm 0.0076$ \\
\hline
$\text{LC}\_\text{m}$ & $0.0867 \pm 0.0065$ \\
\hline
$\text{LC}\_\text{p}$ & $0.0871 \pm 0.0062$ \\
\hline
$\text{OR}\_\text{LSS}\_\text{m}$ & $0.0918 \pm 0.0074$ \\
\hline
$\text{OR}\_\text{LSS}\_\text{p}$ & $0.0877 \pm 0.0061$ \\
\hline
$\text{w}\_\text{m}$ & $0.0874 \pm 0.0063$ \\
\hline
$\text{w}\_\text{p}$ & $0.1050 \pm 0.0086$ \\
\hline
\end{tabular}
}
\label{tab:red}
\end{table}

\section{Void Probability Function in Galaxies}\label{vpf:g}

\subsection{VPF dependence on galaxy morphology}

\begin{figure}[htbp]
\centering
\includegraphics[width=.325\textwidth]{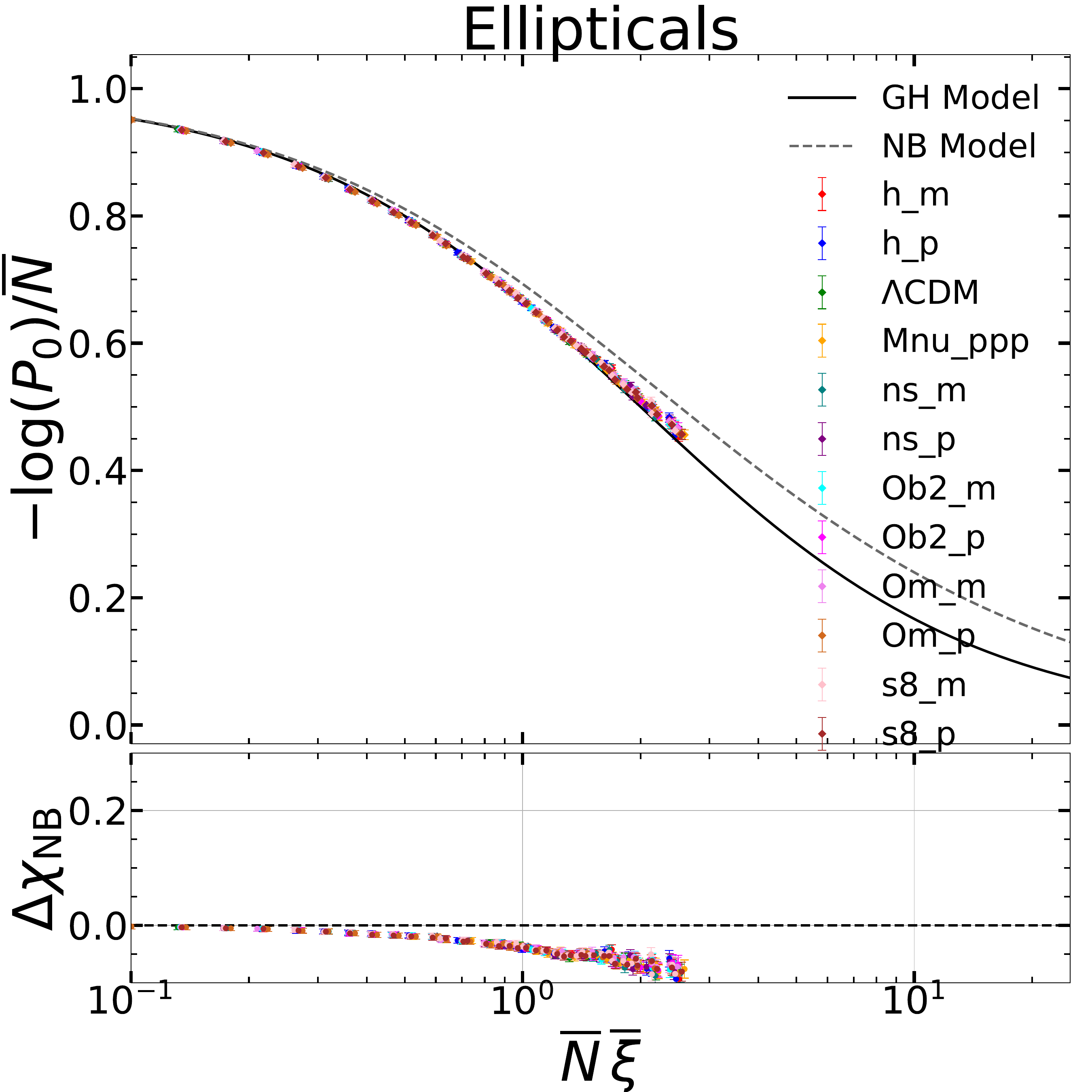}
\includegraphics[width=.325\textwidth]{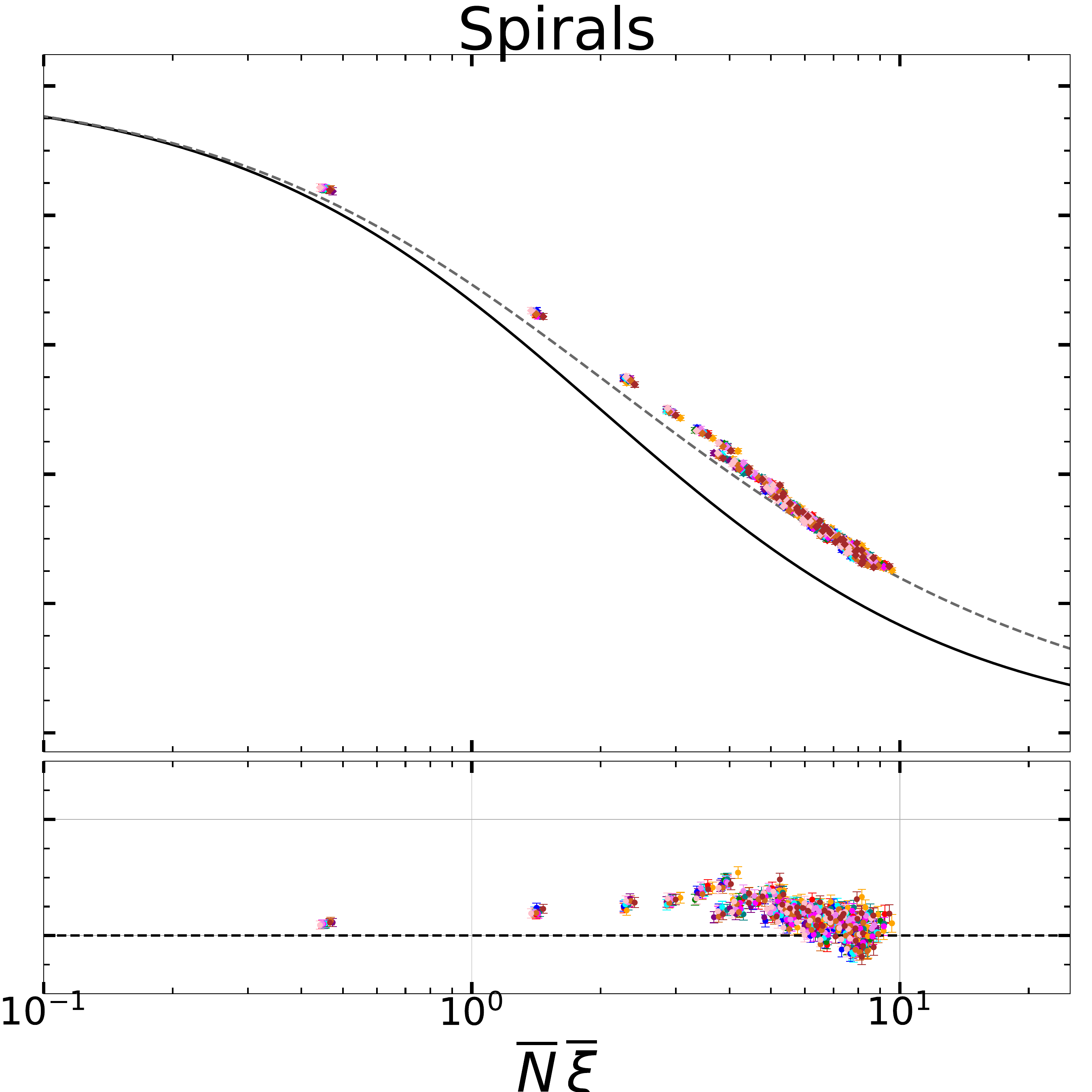}
\includegraphics[width=.325\textwidth]{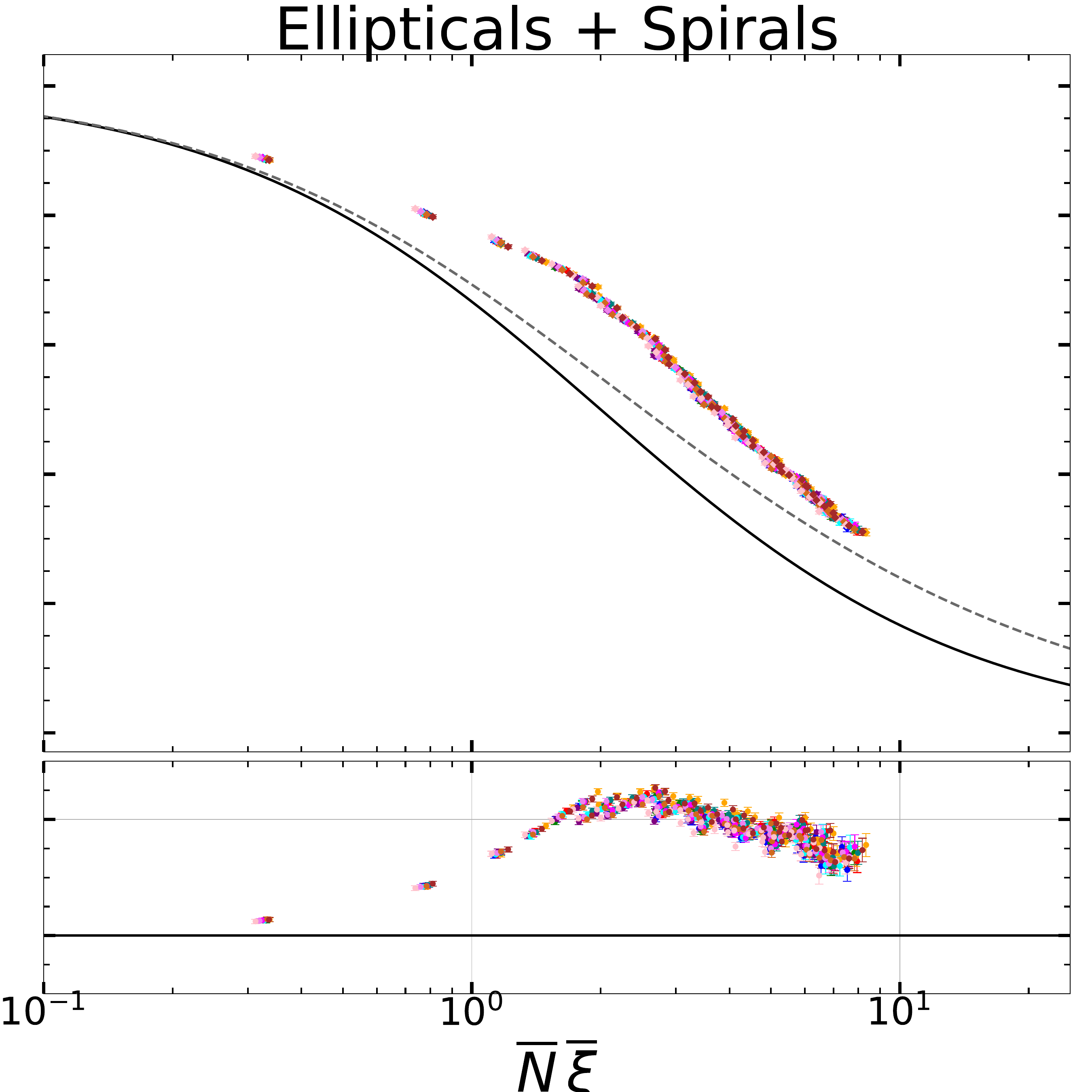}
\caption{VPF of ellipticals (left), spirals (middle), and the combined sample (right). The solid line represents the analytical estimation of $\chi = -\log P_{0}/\overline{N}$ with the GH model, while the dashed line represents the same but for the NB model. The colored markers represent the $\chi$ obtained for the various galaxy catalogues listed in Table~\ref{tab:galaxy}. The bottom panel in each figure shows the difference between the $\chi$ obtained in each galaxy catalogue and that of the NB model. \label{gal:mor}}
\end{figure}

Figure~\ref{gal:mor} presents VPF for galaxies, where the left panel shows the VPF for ellipticals, the middle panel for spirals, and the right panel for the combined sample. The lower panels represent the difference between the $\chi$ obtained in each galaxy catalogue and that of the NB model. The corresponding deviations from the NB model are shown in Table~\ref{tab:gal:mor}. We find that for all the cosmological models, ellipticals show a VPF that closely follows the predictions of the GH model. This is expected since they are typically found in high-density regions such as galaxy clusters and filaments, and therefore are efficient in tracing the underlying matter distribution \citep{dressler1980,zehavi2011}. On the other hand, the spirals, which tend to reside in less dense environments, such as the outskirts of clusters or in fields, exhibit a VPF that aligns more closely with the NB model. Interestingly, the right panel, which includes both ellipticals and spirals, shows a VPF that does not agree well with either model, which could be due to the superposition of two populations with different spatial clustering and bias properties.

\begin{table}[ht]
\centering
\caption{VPF dependence on the galaxy morphology.}
\resizebox{0.7\textwidth}{!}
{
\begin{tabular}{|l||c|c|c|}
\hline
\hline
\multicolumn{1}{|c||}{} & \multicolumn{3}{c|}{Deviations from the NB model for different galaxy morphologies}  \\
\cline{2-4}
\multicolumn{1}{|c||}{Cosmology} & Ellipticals & Spirals & All galaxies \\
\hline
\hline
$h\_\text{m}$ & $0.0479 \pm 0.0025$ & $0.0422 \pm 0.0023$ & $0.1818 \pm 0.0037$ \\
\hline
$h\_\text{p}$ & $0.0473 \pm 0.0025$ & $0.0374 \pm 0.0021$ & $0.1758 \pm 0.0036$ \\
\hline
$\Lambda$CDM & $0.0471 \pm 0.0024$ & $0.0399 \pm 0.0022$ & $0.1803 \pm 0.0037$ \\
\hline
$\text{Mnu}\_\text{ppp}$ & $0.0483 \pm 0.0024$ & $0.0467 \pm 0.0024$ & $0.1903 \pm 0.0038$ \\
\hline
$\text{ns}\_\text{m}$ & $0.0484 \pm 0.0024$ & $0.0418 \pm 0.0022$ & $0.1842 \pm 0.0037$ \\
\hline
$\text{ns}\_\text{p}$ & $0.0458 \pm 0.0024$ & $0.0372 \pm 0.0021$ & $0.1760 \pm 0.0037$ \\
\hline
$\text{Ob2}\_\text{m}$ & $0.0449 \pm 0.0023$ & $0.0405 \pm 0.0022$ & $0.1793 \pm 0.0037$ \\
\hline
$\text{Ob2}\_\text{p}$ & $0.0457 \pm 0.0023$ & $0.0383 \pm 0.0021$ & $0.1804 \pm 0.0038$ \\
\hline
$\text{Om}\_\text{m}$ & $0.0438 \pm 0.0023$ & $0.0421 \pm 0.0023$ & $0.1811 \pm 0.0037$ \\
\hline
$\text{Om}\_\text{p}$ & $0.0476 \pm 0.0025$ & $0.0361 \pm 0.0020$ & $0.1766 \pm 0.0036$ \\
\hline
$\text{s8}\_\text{m}$ & $0.0463 \pm 0.0024$ & $0.0398 \pm 0.0022$ & $0.1729 \pm 0.0035$ \\
\hline
$\text{s8}\_\text{p}$ & $0.0471 \pm 0.0024$ & $0.0445 \pm 0.0023$ & $0.1870 \pm 0.0038$ \\
\hline
\end{tabular}
}
\label{tab:gal:mor}
\end{table}

\subsection{VPF dependence on galaxy subsampling}

\begin{figure}[htbp]
\centering
\includegraphics[width=.32\textwidth]{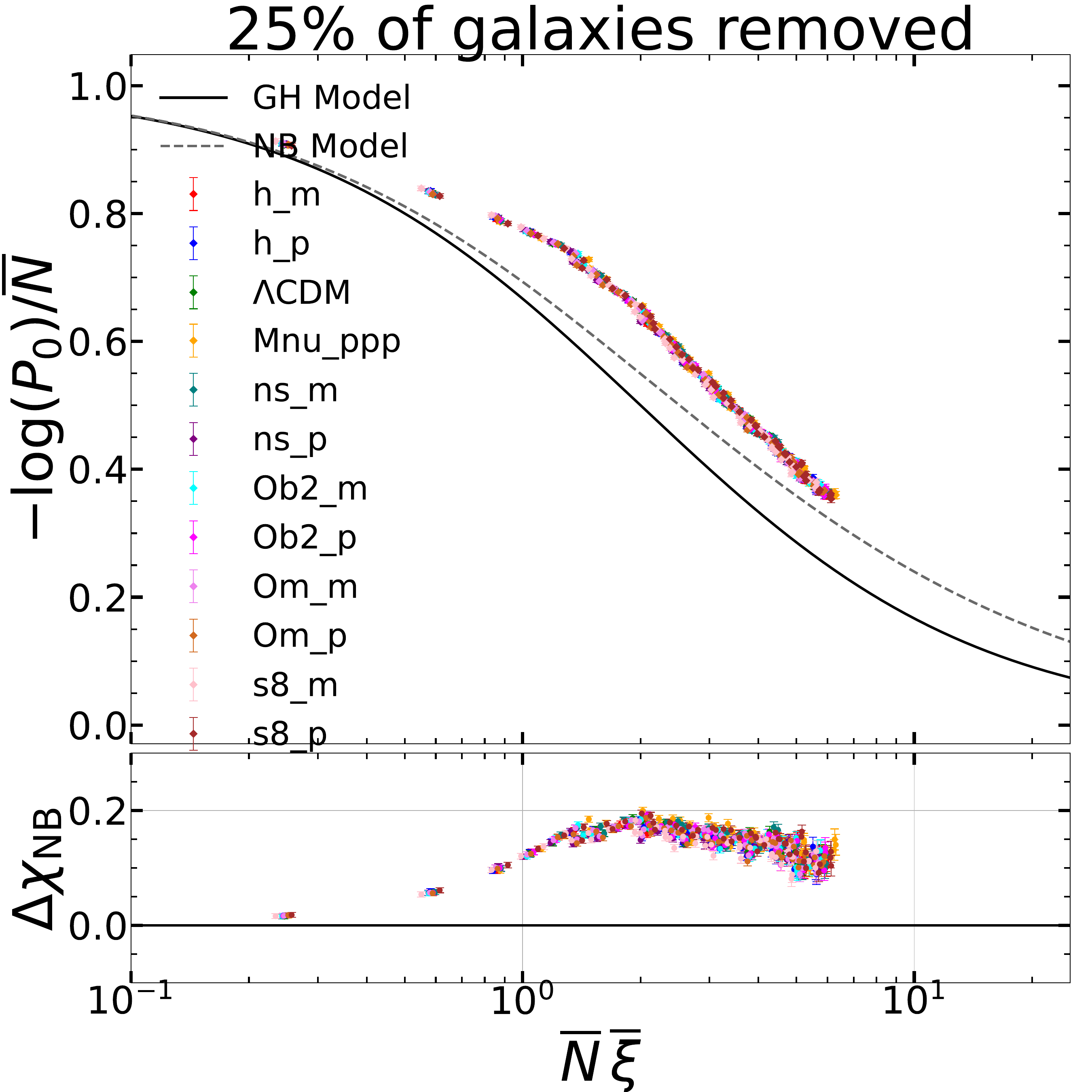}
\includegraphics[width=.32\textwidth]{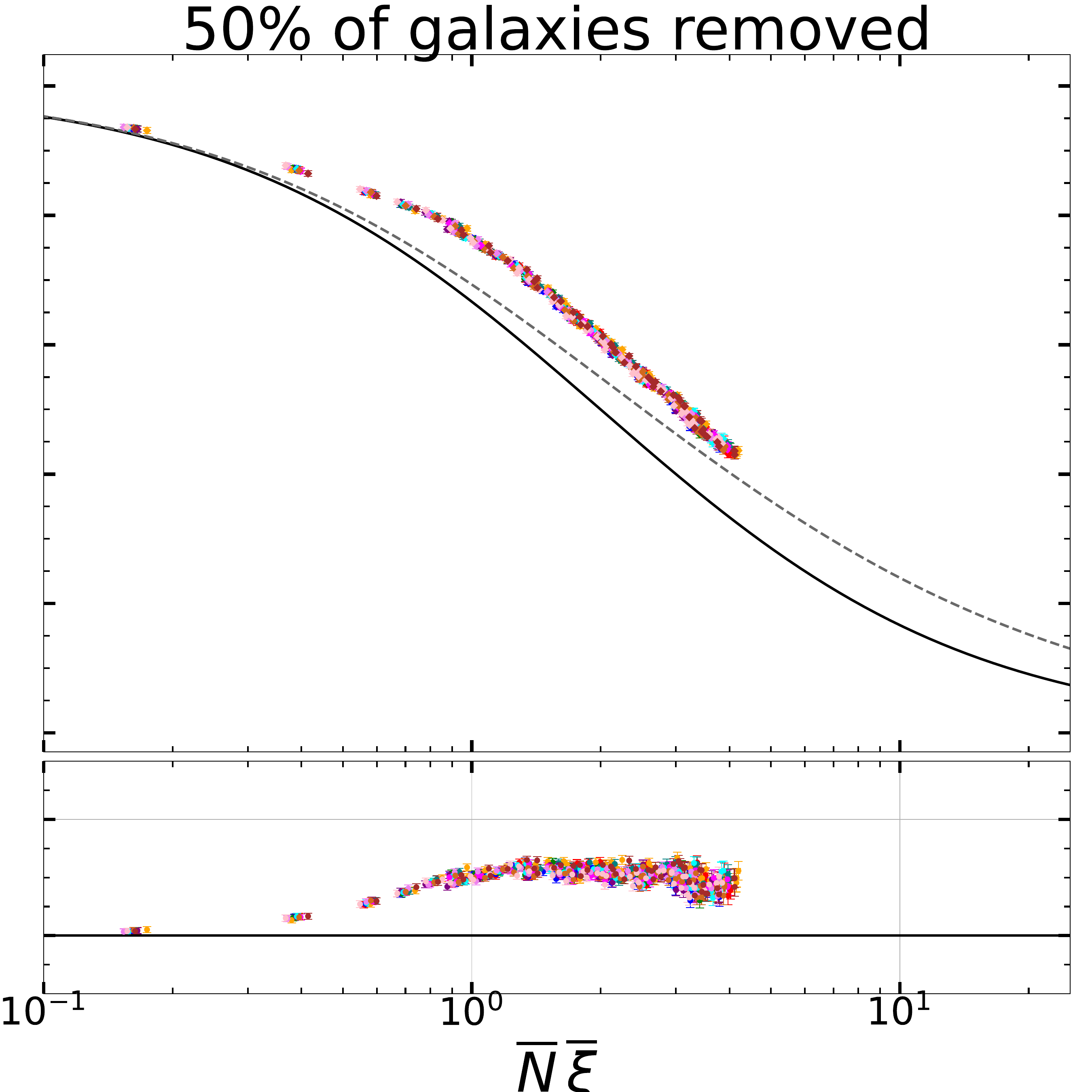}
\includegraphics[width=.32\textwidth]{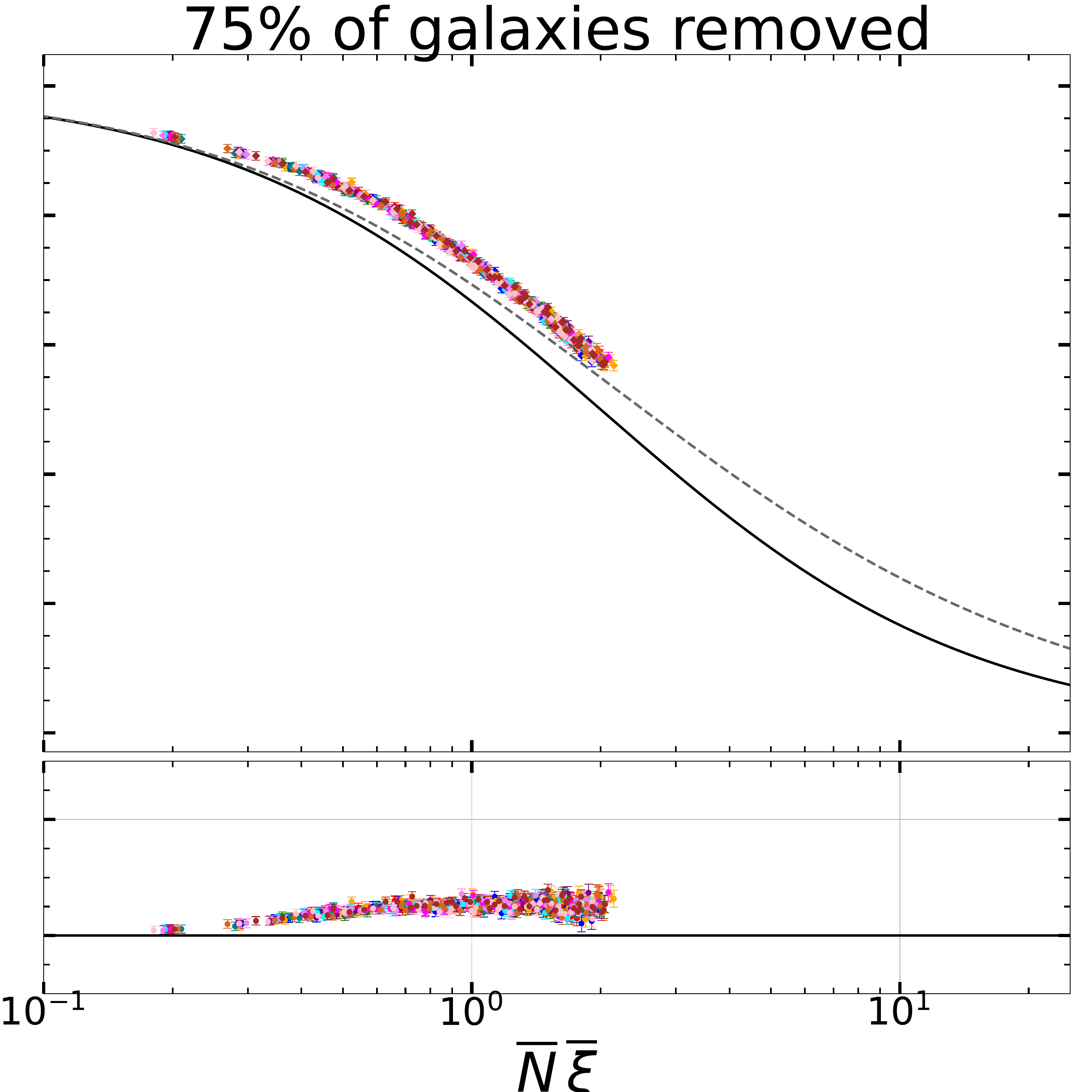}
\caption{Same as Figure~\ref{gal:mor} but for the galaxies with random subsampling by $25\%$ (left), $50\%$ (center) and $75\%$ (right). \label{gal:sub}}
\end{figure}

Figure~\ref {gal:sub} shows the VPF measurements from galaxies after randomly removing $25\%$ (left), $50\%$ (middle), and $75\%$ (right) of the sample. The corresponding deviations from the NB model are listed in Table~\ref{tab:gal:sub}. We find that as the degree of random subsampling increases, the overall discrepancy between the measured VPF and the theoretical models decreases. The trend holds across all cosmological models, with only minor differences in the level of deviation. The observed trend is likely a consequence of random dilution suppressing small-scale clustering, causing the void statistics to be driven by the one-point count distribution. In this regime, the VPF has been reported to be well described by the NB model, which accounts for the over-dispersion of galaxy counts \cite{Fry1986,sheth1996,conroy2005}. Recent work applying similar dilution tests to simulations also finds that the VPF remains well described by the NB model \cite{kurban2023}.

\begin{table}[ht]
\centering
\caption{VPF dependence on the galaxy subsampling.}
\resizebox{0.85\textwidth}{!}
{
\begin{tabular}{|l||c|c|c|}
\hline
\hline
\multicolumn{1}{|c||}{} & \multicolumn{3}{c|}{Deviations from the NB model for different galaxy subsamples}  \\
\cline{2-4}
\multicolumn{1}{|c||}{Cosmology} & $25\%$ galaxies removed & $50\%$ galaxies removed & $75\%$ galaxies removed \\
\hline
\hline
$h\_\text{m}$ & $0.1439 \pm 0.0033$ & $0.1012 \pm 0.0029$ & $0.0506 \pm 0.0023$ \\
\hline
$h\_\text{p}$ & $0.1390 \pm 0.0032$ & $0.0981 \pm 0.0028$ & $0.0462 \pm 0.0022$ \\
\hline
$\Lambda$CDM & $0.1432 \pm 0.0033$ & $0.1002 \pm 0.0029$ & $0.0497 \pm 0.0023$ \\
\hline
$\text{Mnu}\_\text{ppp}$ & $0.1529 \pm 0.0035$ & $0.1062 \pm 0.0030$ & $0.0549 \pm 0.0025$ \\
\hline
$\text{ns}\_\text{m}$ & $0.1462 \pm 0.0033$ & $0.1048 \pm 0.0030$ & $0.0533 \pm 0.0024$ \\
\hline
$\text{ns}\_\text{p}$ & $0.1397 \pm 0.0033$ & $0.0974 \pm 0.0029$ & $0.0473 \pm 0.0022$ \\
\hline
$\text{Ob2}\_\text{m}$ & $0.1419 \pm 0.0033$ & $0.0978 \pm 0.0028$ & $0.0490 \pm 0.0023$ \\
\hline
$\text{Ob2}\_\text{p}$ & $0.1427 \pm 0.0033$ & $0.1014 \pm 0.0029$ & $0.0504 \pm 0.0023$ \\
\hline
$\text{Om}\_\text{m}$ & $0.1430 \pm 0.0033$ & $0.1005 \pm 0.0029$ & $0.0486 \pm 0.0023$ \\
\hline
$\text{Om}\_\text{p}$ & $0.1396 \pm 0.0033$ & $0.0995 \pm 0.0028$ & $0.0492 \pm 0.0023$ \\
\hline
$\text{s8}\_\text{m}$ & $0.1360 \pm 0.0032$ & $0.0954 \pm 0.0028$ & $0.0471 \pm 0.0023$ \\
\hline
$\text{s8}\_\text{p}$ & $0.1494 \pm 0.0034$ & $0.1045 \pm 0.0029$ & $0.0525 \pm 0.0023$ \\
\hline
\end{tabular}
}
\label{tab:gal:sub}
\end{table}

\section{Conclusions}\label{con}

In this work, we analyzed the VPF of dark matter haloes and galaxies using the \textsc{Quijote} N-body simulations across a wide range of cosmological models, including both the standard $\Lambda$CDM scenario and several well-motivated extensions. We examined in detail how the VPF responds to variations in cosmology, as well as to redshift, halo mass, galaxy morphology, and tracer subsampling, and compared the results with theoretical models of hierarchical clustering. For the halo catalogues, we focus our analysis on the influence of the amplitudes of primordial non-Gaussianity, total neutrino mass, the background density contrast parameter, the dark energy parameter, and the $f(R)$ modified gravity, while for the galaxy catalogues, we focus on the variations of the five vanilla $\Lambda$CDM parameters. We note that the galaxy catalogues used in this work are constructed using an HOD prescription, rather than being drawn from full hydrodynamical simulations or semi-analytic models. As a result, baryonic processes are not modelled with high precision. However, the VPF is primarily sensitive to large-scale clustering and number density, and hence we expect our main conclusions to be largely insensitive to these effects. Our results are summarized as follows:

\begin{itemize}
    
    \item At $z = 0$, the VPF for dark matter haloes is well described by the GH model (Figure~\ref{halo:z}), with the $\Lambda$CDM cosmology showing the smallest deviations. In contrast, the scenario incorporating massive neutrinos exhibits the largest deviation, likely due to neutrino free-streaming suppressing small-scale clustering and producing a smoother matter distribution (see \citep{Lesgourgues2006,Navarro2015,Castorina2015}). At higher redshifts, the VPF gradually departs from the GH model and trends toward the NB model (Table~\ref{tab:z_halo}).

    \item The VPF exhibits a dependency on the halo mass (Figure~\ref{halo:mass}), with deviations from the GH model varying across mass bins and cosmologies (Table~\ref{tab:halo_mass}). Low-mass haloes show the largest deviations in the $\Lambda$CDM cosmology, while intermediate- and high-mass haloes show the strongest deviations in the modified gravity scenario. This trend highlights that the mass-dependent clustering of haloes is sensitive to the underlying cosmology, with modified gravity models producing the most pronounced deviations for more massive haloes.

     \item The VPF is sensitive to random subsampling (Figure~\ref{halo:sub}), with higher levels of subsampling producing substantial deviations from the GH model in cosmologies with modified gravity, massive neutrinos, and primordial non-Gaussianity, showing the differences in halo clustering induced by these models more clearly.
    
    \item The VPF shows distinct behaviour in real and redshift space (Figure~\ref{halo:red}), following the GH model in real space, while shifting toward the NB model in redshift space due to the impact of peculiar velocities. Cosmologies with primordial non-Gaussianity exhibit the largest deviations in redshift space, suggesting that the effects of redshift-space distortions are more pronounced in cosmological scenarios that incorporate primordial non-Gaussianities, and could be linked to their amplitudes.

    \item The VPF for galaxies shows a strong dependence on morphology (Figure~\ref{gal:mor}), with ellipticals closely following the GH model across all cosmologies, likely because they are typically found in dense environments such as galaxy clusters and filaments, while spirals are better described by the NB model, consistent with their preference for lower-density regions. Their combined VPF deviates from both models, likely due to the superposition of populations with distinct clustering and bias properties.

    \item Similar to the dark matter haloes, the VPF for galaxies also changes with sample dilution (Figure~\ref {gal:sub}). We find that as the level of random subsampling increases, the VPF progressively approaches the NB model for all cosmologies.

\end{itemize}

The results, therefore, indicate that the VPF and its scaling behaviour provide a useful complementary probe of cosmology, while also being sensitive to the properties of the tracers. The observed dependence of the VPF on specific cosmological parameters suggests that void-based statistics could help break parameter degeneracies and test extensions of the standard $\Lambda$CDM model. This further motivates their use in upcoming large-volume galaxy surveys to obtain additional insights and tighter cosmological constraints. However, applying this method to observational catalogues may require careful assessment of cosmic variance, which introduces a non-negligible scatter and additional uncertainties associated with the measurement of tracer positions.

\acknowledgments

We thank Francisco Antonio Villaescusa Navarro and Marco Baldi for sharing the $f(R)$ cosmology dataset. We also thank the anonymous referee for their valuable feedback, which helped strengthen the manuscript.




\bibliographystyle{JHEP}
\bibliography{biblio}

\appendix  

\subsection{Uncertainties}\label{un}

To understand the agreement between the measurements and the model, we compute
the residuals, the root-mean-square (RMS) deviation, and their associated
uncertainties as follows:
Let $\chi_i$ denote the measured value of the statistic in bin $i$
with uncertainty $\Delta \chi_i$, and let $M_i$ be the model prediction in the
same bin. 
We define the residual in each bin as the fractional deviation from the model as follows:
\begin{equation}
    r_i = \frac{\chi_i}{M_i} - 1,
\end{equation}
where $r_i = 0$ corresponds to a perfect agreement with the model.

The uncertainty on each residual is obtained by the standard error propagation and can be written as:
\begin{equation}
    \Delta r_i = \frac{\Delta \chi_i}{M_i}.
\end{equation}
The overall deviation between the measurements and the model can be summarised by the RMS deviation:
\begin{equation}
    \mathrm{RMS}
    = \sqrt{ \frac{1}{N} \sum_{i=1}^{N} r_i^{2} },
\end{equation}
where $N$ is the total number of bins.
To compute the uncertainty on the RMS deviation, we propagate the uncertainties in the
individual residuals by taking a derivative with respect to $r_i$ and can be written as:
\begin{equation}
    \frac{\partial\,\mathrm{RMS}}{\partial r_i}
    = \frac{r_i}{N\,\mathrm{RMS}}.
\end{equation}
Thus, the variance in the RMS deviation can be written as
\begin{equation}
    \Delta \mathrm{RMS}^{2}
    = \sum_{i=1}^{N}
      \left( \frac{r_i}{N\,\mathrm{RMS}} \Delta r_i \right)^{2},
\end{equation}
which can be condensed into the following form:
\begin{equation}
    \Delta \mathrm{RMS}
    = \sqrt{
        \frac{
            \sum_{i=1}^{N} ( r_i \Delta r_i )^{2}
        }{
            N\,\mathrm{RMS}
        }
      }.
\end{equation}

\subsection{Impact of Cosmic Variance on the VPF}\label{cv}

\begin{figure}[htbp]
\centering
\includegraphics[width=0.6\textwidth]{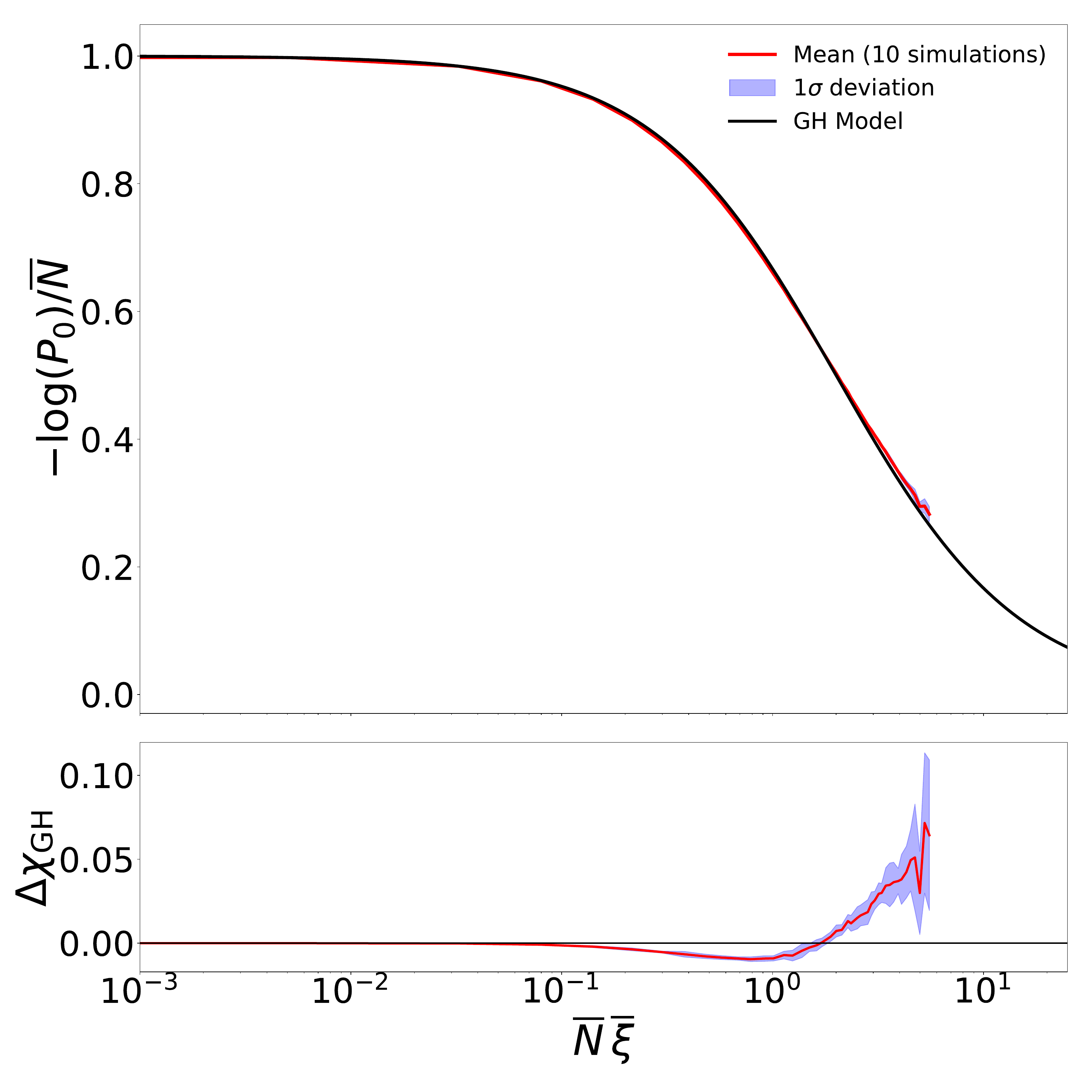}
\caption{VPF of the dark matter haloes for $10$ independent $\Lambda$CDM realizations at $z=0$ using the same halo selection criteria as in Figure~\ref{halo:z}. The mean VPF across realizations is shown in red, while the shaded blue region represents the $1\sigma$ scatter, providing an estimate of cosmic variance. The bottom panel shows the difference between the $\chi$ obtained from the $\Lambda$CDM realizations and that of the GH model. \label{lcdm:ten}}
\end{figure}

To estimate the impact of cosmic variance, we compute the VPF using $10$ independent realizations of the $\Lambda$CDM model at $z=0$. Figure~\ref{lcdm:ten} shows the mean VPF and the corresponding $1\sigma$ deviations, providing an estimate of cosmic variance. For the $\Lambda$CDM model at $z=0$, the mean value obtained from $10$ realizations is $0.031 \pm 0.017$, compared to $0.069 \pm 0.034$ from the single realization reported in Table~\ref{tab:z_halo}. The results indicate that cosmic variance introduces a non-negligible scatter and should be considered when interpreting differences in the VPF across cosmological models and observations.

\end{document}